\documentclass[twocolumn]{aastex631}

\usepackage{tabularray}

\usepackage{savesym}
\savesymbol{tablenum}
\usepackage{siunitx}
\restoresymbol{SIX}{tablenum}

\newcommand{\msun}{M_\odot}
\newcommand{\mdot}{\dot{m}}

\usepackage{xcolor}

\usepackage{amsmath}
\usepackage{siunitx}
\usepackage{graphicx}
\usepackage{booktabs}
\graphicspath{figures}
\usepackage[normalem]{ulem}

\begin{document}

\title{Simulation-Based Prediction of Black Hole Spectra: From $10M_\odot$ to $10^8 M_\odot$}

\author[0000-0003-0685-3525]{Chris Nagele}
\affiliation{Department of Physics and Astronomy\\
Johns Hopkins University\\
Baltimore MD 21218}

\author[0000-0002-2995-7717]{Julian H. Krolik}
\affiliation{Department of Physics and Astronomy\\
Johns Hopkins University\\
Baltimore MD 21218}

\author[0000-0003-0685-3525]{Rongrong Liu}
\affiliation{Center for Astrophysics \\ Harvard $\&$ Smithsonian \\ 60 Garden St., Cambridge, MA 02138, USA}

\author[0000-0002-8676-425X]{Brooks E. Kinch}
\affiliation{Department of Mechanical Engineering and Applied Mechanics\\
University of Pennsylvania\\
Philadelphia PA 19104}

\author[0000-0002-2942-8399]{Jeremy D. Schnittman }
\affiliation{Gravitational Astrophysics Lab, NASA Goddard Space Flight Center, Greenbelt MD 20771}

\begin{abstract}

It has long been thought that black hole accretion flows are driven by magnetohydrodynamic (MHD) turbulence, and there are now many general relativistic global simulations illustrating the dynamics of this process.  However, many challenges must be overcome in order to predict observed spectra from luminous systems. Ensuring energy conservation, local thermal balance, and local ionization equilibrium, our post-processing method incorporates all the most relevant radiation mechanisms: relativistic Compton scattering, bremsstrahlung, and lines and edges for 30 elements and all their ions. Previous work with this method was restricted to black holes of $10 M_\odot$; here, for the first time, we extend it to $10^8 M_\odot$ and present results for two sub-Eddington accretion rates and black hole spin parameter 0.9. The spectral shape predicted for stellar-mass black holes matches the low-hard state for the lower accretion rate and the steep power law state for the higher accretion rate. For high black hole mass, both accretion rates yield power-law continua from $\sim 0.5 - 50$~keV whose X-ray slopes agree well with observations. For intermediate mass black holes, we find a soft X-ray excess created by inverse Compton scattering of low-energy photons produced in the thermal part of the disk; this mechanism may be relevant to the soft X-ray excess commonly seen in massive black holes. Thus, our results show that standard radiation physics applied to GRMHD simulation data can yield spectra reproducing a number of the observed properties of accreting black holes across the mass spectrum.
\end{abstract}


\section{Introduction} \label{sec:intro}

Astronomers have long been confident that a variety of bright X-ray sources are ultimately powered by black holes (BHs), even while the precise mechanism that converts energy lost by accreting matter to radiation is debated. These sources fall into two groups, separated by many orders of magnitude in mass. The lower mass group, consisting of stellar mass BHs, is observable mostly in the Galaxy, where these BHs siphon gas from a companion star. The luminosity associated with the accreted gas is often bright enough to be detected by space based X-ray observatories, while the periodic modulation of the stellar companion's optical line features' Doppler shifts yield confident estimates of the mass of the compact object, masses in the range $M \in 3 - 20\;\msun$ \citep{Remillard_2006,Done+2007}. 

At the higher mass end, supermassive BHs reside at the centers of galaxies, including our own \citep{Kormendy1995ARA&A..33..581K}. Their masses can be probed by the orbits of nearby stars in our Galaxy \citep{Schodel2002Natur.419..694S,Ghez2008ApJ...689.1044G} and in external galaxies by stellar velocity dispersions close to the black hole \citep{Kormendy2013ARA&A..51..511K}; the correlations between these directly measured masses and the velocity dispersion or mass of the host galaxy's stellar bulge provide another means of supermassive black hole mass estimation when it is impossible to obtain the mass directly \citep{Kormendy2013ARA&A..51..511K,McConnell2013ApJ...764..184M}. A subclass of supermassive black holes (active galactic nuclei, or AGN) accrete enough matter from their host galaxy's interstellar medium to make them bright enough to be observed, in some cases even at high redshift. For AGN where a bulge dispersion is difficult to measure, an estimate can be made based on the AGN luminosity. Specifically, if AGN luminosities, typically $\sim 10^{44} - 10^{47}$~erg~s$^{-1}$, are Eddington-limited, they demand black hole masses $\gtrsim 10^6 - 10^9 M_\odot$. 

In this paper, we will explore how the spectrum emanating from the gas near an accreting black hole depends on the mass of the black hole. In general relativity, two black holes with different masses are essentially identical; the mass affects only the characteristic lengthscale and light-crossing timescale in the spacetime near each black hole.
The only expected difference between the two mass categories lies in the thermal properties of the accreting gas: classical disk theory suggests $T \propto {\dot m}^{1/4} M^{-1/4}$ \citep{Shakura1973A&A....24..337S,Novikov1973blho.conf..343N},  where $\dot m$ is the accretion rate in Eddington units.  In rough terms, this is consistent with the contrasts between X-ray binary and AGN spectra---with a mass contrast $\sim 10^7$, the thermal portions of X-ray binary spectra indicate a temperature $\sim 100\times$ that of AGN.

In order to understand black hole accretion in more quantitative detail, general relativistic magnetodyhydrodynamical (GRMHD) simulations may be utilized. These simulations require general relativity rather than Newtonian gravity due to the presence of the black hole, and they require three dimensional (3D) MHD to account for the magnetic stresses which transport angular momentum within the accretion disk \citep{Balbus1991ApJ...376..214B,Balbus1998RvMP...70....1B}.
GRMHD simulations of luminous black holes have accounted for radiative cooling with methods of gradually increasing physicality:
ad hoc cooling functions \citep[e.g.][]{Noble_2009}, local cooling tables \citep{NaetheMotta2025arXiv250508855N}, radiation transfer assuming M1 closure \citep{Liska2022ApJ...935L...1L,Roth2022ApJ...933..226R,Fragile+2023,Roth2025arXiv250118040R}, or multi-angle group radiation transfer \citep{White2023ApJ...949..103W,Jiang2025arXiv250509671J,Zhang2025arXiv250602289Z}.

Relating these GRMHD simulations to spectra, however, has seen comparatively little progress.
Post-processing codes using both Monte-Carlo and characteristics methods have been developed, but have not yet been able to incorporate the physics relevant to optically thick accretion flows (for an overview of these methods, see Appendix A). 
Observational studies of how spectra depend on black hole mass exist \citep{Falcke2004A&A...414..895F,Ruan2019ApJ...883...76R,Moravec2022A&A...662A..28M}, but these are hampered by the dearth in observed intermediate mass black holes spanning several orders of magnitude in black hole mass, and by the practically unobservable regions of the ultra-violet spectrum.

The goal of this paper is to show how the spectra of accreting black holes change as a function of black hole mass by applying a full repertory of radiation mechanisms to density, velocity, and cooling rate data drawn from GRMHD simulations.  
In comparison to previous studies and to other post-processing codes (Appendix A), we include a far more complete set of mechanisms determining ionization balance, thermal balance, and radiation transfer (briefly summarized below) than any previous scheme to predict spectra of accreting black holes. Our method solves the 3D radiation transfer problem in the corona with a Monte Carlo code, \texttt{Pandurata} \citep{Schnittman_2013}, coupling it to a Feautrier radiation solver in the optically thick disk, \texttt{PTransX}; the latter code solves for the local ionization state and temperature \citep{Kinch_2019}. 
The two calculations (each in 3D) are made mutually consistent by having each code provide the radiation boundary conditions for the other at every point on the disk photosphere.
This division of labor allows us to solve for local thermal balance in the vast majority of the system (in particular, where the luminosity is generated) with a method appropriate to the dominant mechanisms, something other post-processing codes are not able to do. This is a crucial step because the nature of the seed photons produced in the disk and then upscattered by inverse Compton scattering in the corona can have a critical impact on the
emergent spectrum.
We use a multi-frequency energy grid spanning seven orders of magnitude in energy color{blue} so as to cover the entire relevant frequency range.
This wide-band frequency grid has sufficiently fine binning to propagate accurately photoionization features, such as Fe~$K\alpha$ lines, something other post-processing codes cannot accomplish. We are preparing a companion paper which describes these lines in detail.

In this paper, we apply this procedure to cases with eight different black hole masses ($M = 10^1, \, 10^2, \ldots 10^8 M_\odot$), two accretion rates ($\dot m = 0.01$, 0.1), and a single spin parameter (0.9).
We will show the simple temperature scaling is present in some parts of the system, but not in others.  In addition, our models successfully reproduce several spectral features of black hole accretion for both low- and high-mass black holes.

The reason that we have developed this complex post-processing procedure is to be able to compare a physically grounded spectrum to observations, and the community does not lack for X-ray observations of accreting black hole systems.  The coincident operation of Chandra \citep{Weisskopf2002PASP..114....1W}, XMM-Newton \citep{Struder2001A&A...365L..18S}, Swift \citep{Gehrels2004ApJ...611.1005G}, NuSTAR \citep{Harrison2013ApJ...770..103H}, NICER \citep{Gendreau2016SPIE.9905E..1HG}, eROSITA \citep{Predehl2021A&A...647A...1P}, IXPE \citep{Soffitta2021AJ....162..208S} XRISM \citep{XRISM2020arXiv200304962X}, and others have allowed broadband X-ray spectral information, polarization degree, and variability to be studied. 
although most of the results of this paper relate to X-ray spectra, our method is suitable at longer wavelengths, as we compute ionization balance solutions which are also relevant for optical and ultraviolet spectra of AGN. The limiting factor for producing a realistic optical spectrum is achieving accretion-disk inflow equilibrium in a simulation over a very wide range of radii,
and we will leave this for future work.

This paper is organized as follows: in Sec. \ref{sec:methods} we describe our workflow and then the specific codes (\texttt{HARM3D}, \texttt{Pandurata}, \texttt{PTransX}) and simulations used in this paper. Then, in Sec. \ref{sec:results}, we describe the thermal properties of our converged solutions as well as the emergent spectrum. In Sec. \ref{sec:discussion}, we compare these spectra to black hole observations, and we conclude in Sec. \ref{sec:conclusion}.

\section{Methods} \label{sec:methods}

Our production of X-ray spectra from accreting black hole systems can be divided into two steps. First, GRMHD simulations of the black hole and accretion disk must be run for a sufficient length of time so that the region of interest in the accretion disk has reached inflow equilibrium; otherwise, the disk's surface density profile will reflect the simulation's initial conditions more than the physical mechanisms of disk evolution (Sec. \ref{sec:methods_HARM3D}). Next, individual snapshots are post-processed using the radiative transfer codes briefly described in the Introduction; we extend that description here (Secs. \ref{sec:methods_Pandurata}, \ref{sec:methods_PTransX}).

\subsection{\texttt{HARM3D}} \label{sec:methods_HARM3D}

\begin{figure*}
\centering
\includegraphics[width=\linewidth]{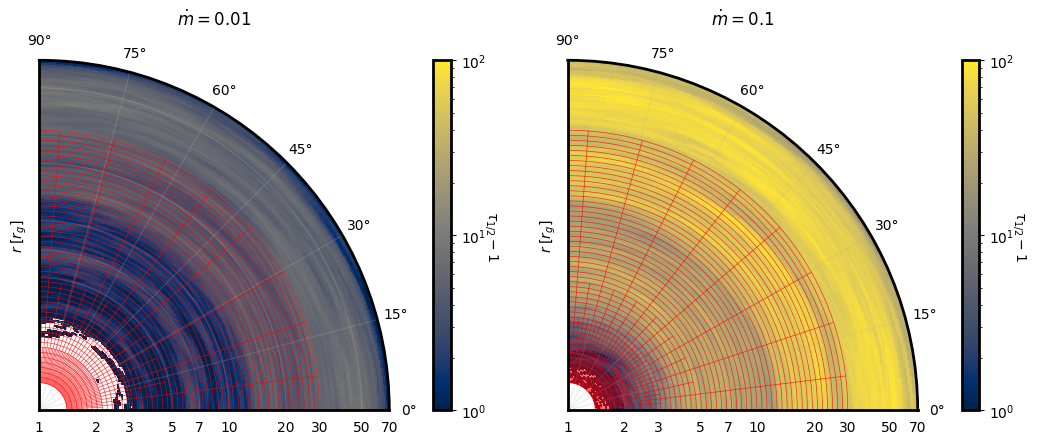}
    \caption{Half optical depth ($\tau$ integrated over $\theta$ from pole to midplane) minus one for each accretion rate case. The white region corresponds to the region where no $\tau=1$ photosphere exists or the area interior to the computational domain.
    The red overlay shows the \texttt{PTransX} grid.
    }
    \label{fig:tau}
\end{figure*}

All the simulations reported here used \texttt{HARM3D} \citep{Noble_2009}, which solves the GRMHD equations in flux-conservative form in three spatial dimensions. The underlying spacetime is that of a Kerr black hole, although more general metrics, including dynamical spacetimes, are possible.
The accretion disk is initialized as a hydrostatic torus like that of \citet{DeVilliers2003ApJ...599.1238D}, threaded by magnetic fields lying along contours of constant density. Built into \texttt{HARM3D} is a cooling function which cools the gas when it goes above a target temperature or entropy. This cooling enforces a fixed aspect ratio H/R; for the simulations we use here, $H/R= 0.06$, appropriate for the accretion rates under consideration.

The two snapshots analyzed in this paper are taken from new simulations that began with late-time data from the spin 0.9 \texttt{HARM3D} simulation published in \citet{Noble_2009}. Both simulations cover a quarter sphere with radial coverage from just inside the horizon out to $70\;r_g$ (Fig. \ref{fig:tau}). The original runs were scale-free with respect to black hole mass because the units of length and time are $GM/c^2$ and $GM/c^3$, respectively. They also are scale-free in terms of mass density. They were extended by new runs in which the unit of mass surface-density was defined by a specific choice for $\mdot$, permitting identification of the Compton scattering photosphere:
\begin{equation}
\label{Eq:rho_cgs}
\rho_{\text {cgs }} r_g =\rho_{\text {code }} \frac{4 \pi }{\kappa} \frac{\mdot / \eta}{\mdot_{\text {code }}}.
\end{equation}
Here $r_g=GM/c^2$ plays the role of the code-unit for length; it can be freely reinterpreted in terms of physical units by choosing $M$ after the simulation has been run. An analogous definition is used to set the luminosity density scale:
\begin{equation}
\label{Eq:cgs}
\mathcal{L}_{\text {cgs }} =\mathcal{L}_{\text {code }} \frac{4 \pi c^3}{\kappa r_g^2} \frac{\mdot / \eta}{\mdot_{\text {code }}}
\end{equation}
where $\mdot = \dot{M}/\dot{M}_{\rm Edd}$ is the accretion rate in Eddington units, $\kappa = \SI{0.4}{cm^2 g^{-1}}$ is the electron scattering opacity, $\dot{m}_{\rm code}$ is the mass accretion rate in code units and $\eta =0.1558$ is the Novikov-Thorne efficiency for an $a = 0.9$ black hole. 

Supported by a choice of $\dot m$, the location of the photosphere can be determined. Fig.~\ref{fig:tau} shows the half optical depth minus one (half optical depth being the Thomson optical depth from the pole to the midplane) as a function of radius and azimuth. The white regions of the plot are where the density is sufficiently low that no $\tau=1$ photosphere exists. Outside the photosphere, the original simple cooling prescription is replaced by one approximating the Compton cooling due to upscattering of thermal photons emitted by the disk \citep{Kinch_2020}.
We have made two new runs, one each for $\mdot  = 0.01$ and $\mdot  = 0.1$. The former was run for $6000M$ past the starting point \citep{Liu2024arXiv241201984L}. From the seven snapshots presented in \citep{Liu2024arXiv241201984L}, we chose $t \simeq 4000 M$, which had a luminosity closest to the nominal accretion rate. For the $\mdot = 0.1$ case, we chose a snapshot at $t \simeq 3000M$.  Because most of the cooling is generated in the corona, where the dynamical and thermal times are short, we do not expect the lightcurves for the two accretion rate cases to display similar time variation, even though they started from the same \texttt{HARM3D} snapshot. 

\subsection{From GRMHD to post-processing}

The main body of information transferred from \texttt{HARM3D} is one or more snapshots; that is, 3D maps at one or more times showing the spatial distribution of the proper rest mass density of the gas, 
the spatial components of the gas's 4-velocity, and the fluid-frame cooling rate.
Although the spin parameter and accretion rate are input parameters to the \texttt{HARM3D} simulation, several additional parameters must be specified before the post-processing is begun. These parameters include the black hole mass, the abundances of 30 elements
(these are important to \texttt{PTransX}, but generally not to Pandurata), and the region of the simulation in which to compute seed spectra using \texttt{PTransX}. Unlike in previous papers, we now compute \texttt{PTransX} only where inflow equilibrium has been achieved, and $r_{\rm inflow} = 30 \; r_g$ for these simulations.  Note that the outer radial bound for \texttt{Pandurata}'s Monte Carlo transfer solution is $r_{\rm outer} = 70 \; r_g$.

\subsection{\texttt{Pandurata}} \label{sec:methods_Pandurata}

\texttt{Pandurata} \citep{Schnittman_2013b} is a Monte Carlo ray-tracing code for application to accretion flows that include both optically thick disks and optically thin coronae in the vicinity of a Kerr black hole. Photon packets are launched from a photosphere and travel along geodesics until possibly scattering off an electron in the corona. Klein-Nishina scattering was implemented in \citet{Kinch_2019}, but pair physics is not included.
After each \texttt{Pandurata} run has traced enough packets to keep statistical fluctuations below a tolerable level, each cell's total energy transfer from gas to photons is computed and compared to the approximate Compton cooling rate in that cell from the underlying GRMHD simulation.  Each cell's electron temperature is then adjusted, via the Newton-Raphson method, so that the Compton cooling rate recorded by the \texttt{Pandurata} run matches that of the GRMHD simulation.  Additional \texttt{Pandurata} runs are then performed until each cell is brought into thermal balance. Unlike in previous papers, we now freeze the temperature of cells that have converged to the desired power to within a factor of one half of the convergence criterion. This step is necessary for higher accretion rate cases where cells near the black hole have a large number of scattering events involving previously-scattered photon packets. The power dissipated in these cells depends not only on the temperature of the cells, but also on the temperature of the cells that hosted the previous scattering events for the relevant photon packets. In extreme cases, the Newton-Raphson algorithm has no chance of converging because of large power fluctuations independent of local temperature. By freezing cells matching the desired power well, we eventually damp those power fluctuations in the problematic cells, allowing them to converge. 

To initialize \texttt{Pandurata}, we begin with a simulation snapshot from \texttt{HARM3D} in which the data are cell-centered on the simulation's spherical grid. We then interpolate the density, temperature, and velocity data to a different spherical grid made suitable for \texttt{Pandurata} principally by coarsening its radial resolution. Thus, while \texttt{HARM3D} has 1020 logarithmically spaced radial gridpoints in $r=1.4-70\;r_g$, \texttt{Pandurata} has 192 logarithmically spaced radial gridpoints in this range.

To begin a \texttt{Pandurata} radiation transfer solution, it is necessary to specify a boundary condition along the disk's photosphere at which photons are injected outward into the surrounding corona.  In most instances, the seed photons are determined by the most recent \texttt{PTransX} solution.  However, at the start of a post-processing run, there has been no previous \texttt{PTransX} solution; in this situation, at each photospheric area cell, the $\texttt{HARM3D}$ cooling rate is integrated vertically from the disk's top photosphere to its bottom photosphere.  The photospheric spectral surface brightness is then taken to be thermal for a bolometric surface brightness equal to half the integrated cooling rate in the column.

In later iterations, the \texttt{PTransX} solution is used for radii within the inflow-equilibrium radius, but at larger radii we
switch to a thermal spectrum related to the temperature of a classical accretion disk using the Newtonian reduction factor for zero ISCO stress \citep{Shakura1973A&A....24..337S}. If $r_{\rm inflow} \gtrsim 30r_g$, both non-zero ISCO stress and the Newtonian approximation introduce errors at the few tens of percent level \citep{Noble_2010,Krolik1999}.  For the actual seed photon injection rate, we scale the classical surface brightness the disk would have if it were in inflow equilibrium with the designated accretion rate by the ratio of dissipation within the disk photosphere to total dissipation: 
\begin{multline}
    \label{Eq:NT}
        kT_{\rm disk}(r) = \Bigg( \frac{3}{2} \frac{c^3}{r_g \kappa_T \eta} \frac{\dot{m}}{(r/r_g)^3} \left[1-(r_{\rm ISCO}/r)^{1/2}\right] \\ 
            \times \frac{\int_{\rm disk} \int_0^{2\pi} \mathcal{L}_{\text {cgs }} d\phi d\theta}{\int_{0}^\pi \int_0^{2\pi} \mathcal{L}_{\text {cgs }} d\phi d\theta} \Bigg)^{1/4},
\end{multline}
where the subscript ``disk" on the integral denotes integration from the upper photosphere to the lower photosphere. After substitution of the Shakura-Sunyaev surface brightness at radii greater than the inflow-equilibrium radius, we ran several tests to see what effect this would have on the converged spectrum and found it to be minor (Appendix B). In this paper, we opt to compute \texttt{PTransX} out to the inflow-equilibrium radius and \texttt{Pandurata} out to the outer \texttt{HARM3D} radius
on the basis that this choice, although necessarily approximate, gives the most realistic outcome.  Fortunately, the luminosity produced where we assume a modified classical surface brightness is at most a small contributor to the spectrum, even at relatively low photon energies.

\subsection{\texttt{PTransX}} \label{sec:methods_PTransX}

\texttt{PTransX} \citep{Kinch_2016,Kinch_2019,Kinch_2021} is a radiation post-processing code designed for use in optically thick black hole accretion disks.
It receives 3D maps of density and MHD heating (that is, the cooling rate) from the \texttt{HARM3D} simulation, but its radiative boundary condition is the flux incident on the disk photosphere as determined by \texttt{Pandurata}.  \texttt{PTransX} solves the 1D time-steady radiation transfer equation within a vertical (or in our spherical coordinates, approximately vertical) slab using the Feautrier method \citep{Mihalas1985JCoPh..57....1M} for a very large number of photon energies. Fig.~\ref{fig:tau} has a red grid overlaying the color-map, which denotes the boundaries of these slabs. In each slab, temperature and ionization balance (for all the ionization stages of all elements up to and including Zn, $Z=30$) are found using an iterative Newton-Raphson scheme. At each iteration, emissivity and absorption coefficients for Compton scattering, bremsstrahlung, and atomic absorption/emission are computed using XSTAR \citep{Kallman2001ApJS..133..221K,Mendoza2021Atoms...9...12M}.

There are two criteria for slab convergence. The first is that energy balance in the slab as a whole is satisfied to within $5\%$. The second is that two successive temperature iterations, both of which satisfy the energy balance condition, cannot have successive Feautrier solutions which differ by more than a certain amount. In this paper, the latter criterion is that the two solutions' flux traveling out of the photosphere cannot differ by more than $5\%$ for each energy grid-point
within 2 orders of magnitude on either side of the peak energy of the outgoing spectrum.

There are several differences between this version of \texttt{PTransX} and previous versions. These differences are centered around the contrast between slabs with and without a thermal core. In all versions of \texttt{PTransX}, local thermodynamic equilibrium is assumed to apply in the thermal core, so that its influence on the outer regions of the disk can be defined in terms of outgoing flux and the temperature at the boundary of the thermal core.

When a thermal core exists, we describe the slab as ``cleaved"; when one does not exist, it is ``whole". \textit{Cleaved} slabs have three regions, the thermal core plus an \textit{upper} and a \textit{lower} region; \textit{whole} slabs comprise only a single region.  The boundary between the thermal core and the upper/lower regions we refer to as the \textit{cleaving location}.  In a cleaved slab, the upper and lower regions are solved individually by \texttt{PTransX}; in a whole slab, \texttt{PTransX} solves the entire slab as a single problem.

In previous versions, a thermal core was said to exist when the scattering optical depth was above a certain threshold, for instance 10 in \citet{Kinch_2021}. We have modified this criterion so that we cleave a slab when the thermalization optical depth exceeds a threshold
\begin{equation}
\label{Eq:cleaving condition}
    \tau_{\rm therm} = z \sqrt{\alpha_a(\alpha_s+\alpha_a)} > 10.
\end{equation}
Here $z$ is the distance from the photosphere to the midplane, $\alpha_a$ is the absorption coefficient for a photon with energy equal to $kT$ and $\alpha_s$ is the scattering coefficient. $\alpha_a$ has contributions from free-free, free-bound, and bound-bound transitions, so it is necessary to solve for ionization balance to compute it. This computation is done by XSTAR \citep{Kallman2001ApJS..133..221K}.

When Eq.~\ref{Eq:cleaving condition} is satisfied, \texttt{PTransX} generally fails to converge. Thus, the region in which we compute \texttt{PTransX} is defined to be outside the thermal core, where a solution can be found.
This is a reasonable approach because we know the spectrum of the light emerging is blackbody, and its flux must match the dissipation rate within the core.
When a slab is cleaved, we choose the cleaving locations to be the outermost gridpoints where $\tau_{\rm therm}>1$. After testing several different values of the $\tau_{\rm therm}$ cleaving threshold, we chose to cleave at $\tau_{\rm therm} \approx 1$ because this choice is consistent with the interpretation outlined above, allowed for convergence in most slabs, and was computationally less expensive than other choices.

However, this procedure contains a subtle difficulty: there is no good approach by which to compute the thermalization optical depth {\it a priori}, as the absorption coefficient varies strongly with temperature.  To work around this difficulty, we estimate $\tau_{\rm therm}$ from the information at hand (either our standard initial condition, described two paragraphs ahead, or the result of the previous \texttt{PTransX} iteration).  If the estimated $\tau_{\rm therm} < 10$ and a whole slab calculation converges while maintaining that condition, we accept this solution.  If, on the other hand, $\tau_{\rm therm}$ in this or a subsequent iteration rises to $>10$ (most often due to a temperature plunge around the midplane), we cleave the slab at $\tau_{\rm therm} = 1$ and proceed with a cleaved solution.

Because convergence is an iterative process, there is no guarantee that the $\tau_{\rm therm} = 1$ surface determined in a particular iteration will coincide with the thermal surface of the converged solution.
We are thus forced to adjust the cleaving location inward or outward based on the temperature profiles of the cleaved slabs during their own iterations. In concert with these adjustments, we also adjust the cleaving location outwards (toward the scattering photosphere) if at some point during the iteration the slab temperature drops below $10^4$~K. Below this temperature, XSTAR is missing relevant physics, and regions of the disk this cold will almost certainly be thermal.

One of the most important determinants of the temperature profile and the outgoing spectrum is the spectrum of radiation incident at the slab boundaries. When this boundary is the scattering photosphere, the incident spectrum is taken from \texttt{Pandurata}. A cleaved slab, however, has one boundary abutting the thermal core. The core has a certain amount of MHD dissipation within it,
and knowledge of this dissipation amount determines the net flux leaving the core:
\begin{equation}
    \label{Eq:thermal core energy balance}
F_{\rm out} = F_{\rm in} +
    (1/2)\int_{\rm core} \, dz \, \mathcal{L} ,
\end{equation}
where $F_{\rm out}$ is the bolometric flux traveling from the thermal region into the slab and $F_{\rm in}$ is the bolometric flux traveling from the slab into the thermal region. The factor of two is there because flux travels out into both the upper slab and the lower slab. Both $F_{\rm in}$ and $F_{\rm out}$ are {\it a priori} unknown and depend non-trivially on the temperature profile, the dissipation in the slab, and the external illumination. Thus, in order to satisfy Eq.~\ref{Eq:thermal core energy balance}, we use a bisection scheme to determine the amplitude of $F_{\rm out}$ ($F_{\rm in}$ is computed for each temperature iteration using the Feautrier solution).

We give the radiation at this boundary a blackbody spectrum with temperature $T_{\rm core}$. This requires two choices on our part, first to fix the temperature instead of letting it vary between each temperature update, and second the choice fo $T_{\rm core}$ as the temperature of the blackbody spectrum. Regarding the first decision, allowing the temperature defining the boundary spectrum to float with the slab's overall solution interferes with numerical convergence, and the majority of slabs will not converge with a variable boundary spectrum. Regarding the second, we chose the specific temperature at which it is fixed to be $T_{\rm core}$ on the grounds that it is a characteristic temperature scale for this region.
However, this choice may be an underestimate, as the radiation temperature at this location is higher by a factor of two (Fig.~\ref{fig:Tcore}). We plan to investigate this question in more detail in future work.

For whole slabs (i.e., uncleaved slabs), our initial temperature guess is
a uniform temperature: $10^6$~K is the default, but it is raised to the effective temperature of the incident flux if that is $>10^6$~K.
For cleaved slabs, we start by setting the temperature at the boundary of the core to $T_{\rm core}$, the effective temperature associated with the dissipation in the core.
From there, we increase the temperature logarithmically so that the total increase is a factor of 100 over the height of the slab, i.e. $T = T_{\rm core} 10^{2|\tau-\tau_{\rm core}|/|\tau_{\rm phot} - \tau_{\rm core}|}$. Such a dramatic increase is expected in this regime \citep{Rozanska2015A&A...580A..77R}, and our experimentation has found this guess to lead most quickly to convergence over a wide range of physical situations. Even in the case of zero incident flux, the temperature still increases near the photosphere due to the combination of magnetic dissipation and inefficient photon radiation processes in this region.

\subsection{Convergence} \label{sec:methods_convergence}

Each of the radiation transfer codes described in the previous two sections provides the radiation boundary condition for the other at the scattering photosphere. We iterate between the two codes until the observed \texttt{Pandurata} spectrum matches the spectrum from the previous iteration to within $5\%$ everywhere within the energy range $0.01\leq \nu < 100$~keV (note that our total energy range is $0.001\leq \nu < 1000$~keV). It is worth remarking that demanding convergence over such a wide range of energies encompassing many different physical processes highlights both the strengths and challenges of our approach.

Although we have no physically motivated argument for why the iterations between \texttt{Pandurata} and \texttt{PTransX} should in general converge, in our experience convergence is achieved reasonably quickly.
Almost all our cases converged within three ``grand"  iterations, i.e. three complete \texttt{Pandurata}/\texttt{PTransX} solutions. When there are relatively few line features in the final spectrum (as for many of the $\mdot=0.1$ cases), convergence was reached even more rapidly.  During this convergence process, the thermal solutions of successive iterations also become more and more similar, both in \texttt{PTransX} slabs and in the corona.  In the first iteration or two,  tens to hundreds of temperature updates are often required for convergence, but later in the process only a few are needed.
In this regard, the higher mass models are not intrinsically more difficult than the lower mass models, despite having lower disk temperatures with more complex atomic physics. It is sometimes the case that an individual \texttt{PTransX} iteration can take longer for the higher masses (due to more cleaved slabs, and more atomic physics), but this is a secondary effect. Typical convergence times of individual slabs on a single core range from 5 to 120 minutes, and with $\sim500$ slabs across 240 cores, a single \texttt{PTransX} iteration takes of order 1 hour (a single slab will be parallelized across multiple cores if there are inactive cores). In contrast, \texttt{Pandurata} general takes 2-3 hours across 48 cores. A typical number of grand iterations back and forth between the two codes is $\sim$5, so that a single model generally takes up a few thousand core-hours.

\section{Results} \label{sec:results}

\begin{figure*}
\includegraphics[width=\linewidth]{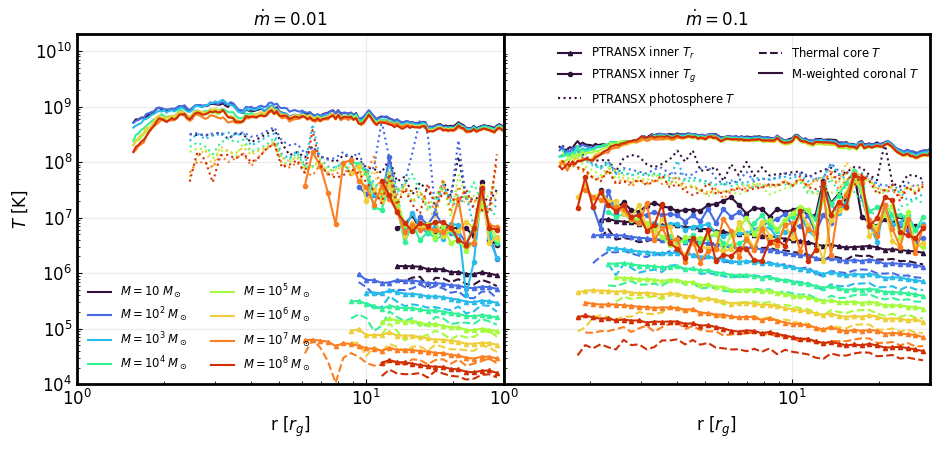}
    \caption{Several temperatures as a function of radius for each mass (color; see legend) and accretion rate (left: $\mdot = 0.01$, right: $\mdot = 0.1$). The dashed lines show the azimuthally averaged thermal core effective temperature.
    The solid lines with triangles show the radiation temperature one cell outside the thermal core (that is, at the base of the atmosphere), while the solid lines with circles show the gas temperature at this location.
    The dotted lines show the photospheric gas temperature. The solid lines show the mass-weighted coronal temperature, averaged over azimuthal and polar angle.}
    \label{fig:Tcore}
\end{figure*}

 This section is divided into three subsections, each dealing with a primary element in our post-processing machinery: local thermal balance in several different subregions, energy exchange between the disk and the corona, and the spectra emerging.

\subsection{Local thermal balance} \label{sec:results_local_thermal_balance}

The dominant effects controlling the local temperature vary in different subregions of the disk and its environment. We will therefore divide our discussion into four parts, each focusing on a specific zone: the thermal core of the disk, the disk atmosphere, the near-photospheric portion of the corona, and the main volume of the corona. This discussion will be guided by Fig. \ref{fig:Tcore}, which shows the radial dependence of the temperatures in these regions.

\subsubsection{The disk body and its thermal core}

The two panels of Fig.~\ref{fig:Tcore} show that $T_{\rm core}$, 
which represents the dissipation within the thermal core, falls from $\sim 10^7$~K to $\sim 10^4$~K as $M$ increases from $10M_\odot$ to $10^8 M_\odot$.
At the order of magnitude level, these numbers are consistent with those estimated on the basis of classical disk theory and share their
scaling $T \propto M^{-1/4}$.  They are somewhat lower than typical estimated values partly because our examples are sub-Eddington ($\mdot = 0.01,\ 0.1$) and partly because the cores contain only a fraction of the total dissipation. 
The radial temperature-dependence is roughly consistent with the classical expectation, in which $T \propto r^{-3/4} R_R^{1/4}$, where, for rapidly-spinning black holes, $R_R \approx 0.4 (r/r_g)\log (r/r_g)$ for $r/r_g \lesssim 30$ and approaches unity as $r/r_g \rightarrow \infty$.  Because the factor $R_R(r)$ is a substantial correction in these inner regions, the effective temperature profile is much flatter than the $r^{-3/4}$ scaling expected at larger radii.  The thermal core temperature also exhibits fluctuations around the classical prediction because the fraction of the total dissipation contained within a core varies substantially from place to place.

\begin{figure*}
\includegraphics[width=\linewidth]{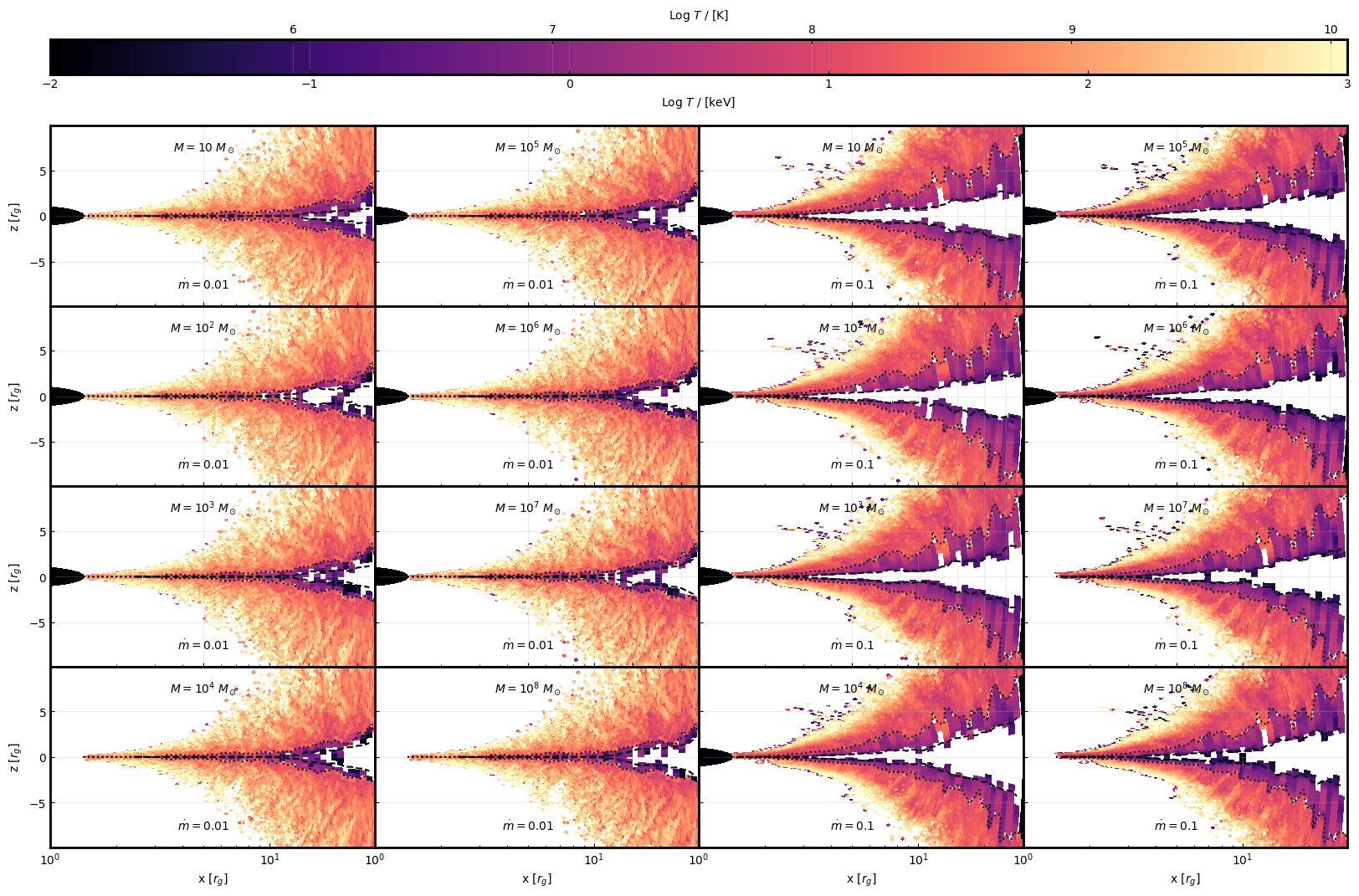}
    \caption{Temperature solutions for an azimuthal slice where each panel shows one model. The black dotted line shows the photosphere in this azimuthal slice, while the black dashed line shows the azimuthally averaged thermal core boundary of \texttt{PTransX}. The region outside the photosphere is the \texttt{Pandurata} solution (note that Fig. \ref{fig:maps} has a different temperature scale) and the region inside the photosphere is the \texttt{PTransX} solution. White regions within the photosphere are either the thermal core (where it exists) or slabs which fail to converge. }
    \label{fig:T}
\end{figure*}

However, the {\it actual} radiation and gas temperatures just outside the thermal core are hotter, by a factor $\sim 2$ for $T_{\rm r}$ and by anywhere from a factor of $\sim 2$ to a factor of $\sim 500$ for $T_{\rm g}$ (see Fig.~\ref{fig:Tcore}). Here we define $T_{\rm r}$ as the temperature for which the LTE radiation energy density matches the actual radiation energy density. $T_{\rm r}$ increases outwards by $1\%-50\%$ over the length of the slab.
Although $T_{\rm r}/T_{\rm core} \approx 2$ almost everywhere, $T_{\rm g}/T_{\rm core}$ varies widely, roughly $\propto M^{1/4}$.
The ratio $T_{\rm g}/T_{\rm core}$ also increases for smaller $\mdot$, in large part because the cores are vertically thinner when $\mdot$ is smaller (see Fig.~\ref{fig:T}), so that they contain a smaller fraction of the total dissipation of the system.

That $T_{\rm g}$ should be so much hotter than $T_{\rm core}$ is easily understood.  Where a thermal core exists ($r \gtrsim 10r_g$ for $\mdot = 0.01$, $r \gtrsim 2r_g$ for $\mdot = 0.1$, with little dependence on $M$: see Fig.~\ref{fig:T}), the disk is quite optically thick: the Thomson optical depth from the photosphere to the thermalization surface ranges from 0.5 to 40 (see Fig.~\ref{fig:tau} for the optical depth all the way to the midplane). As a result,
the inner regions are ``blanketed", i.e., the cooling rate is retarded by the multiple scatters photons must make before escaping.
In addition, the ionizing flux penetrates well below the photosphere, suppressing atomic absorptive opacity (and consequently atomic emissivity), which forces the gas temperature to rise further above the effective temperature.

That $T_{\rm g}$ and $T_{\rm r}$ can be so different from one another in a region just outside the thermalization surface is more surprising. Experimentation with much finer spatial resolution in this region suggests that this is a numerical artifact stemming from a very sharp gradient in $T_{\rm g}$ immediately outside the thermalization surface.  Unfortunately, the resolution required to resolve this region is too costly to employ in every such instance.

\begin{figure}
\includegraphics[width=\linewidth]{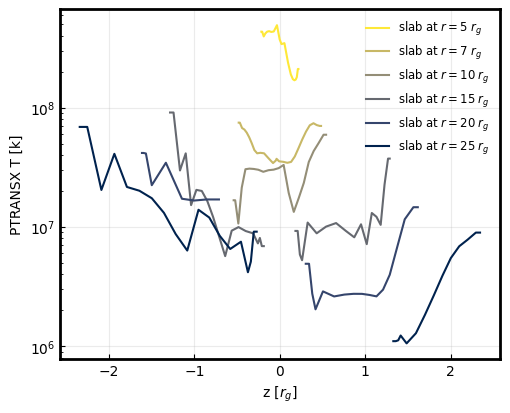}
    \caption{Several example temperature \texttt{PTransX} temperature solutions for $\phi=0$ in the $\mdot=0.01$, $M=10\;\msun$ model (corresponding to the upper left panel in Fig. \ref{fig:T}).}
    \label{fig:Texamples}
\end{figure}

\subsubsection{The disk atmosphere}

For the purposes of this paper, we define the disk ``atmosphere" as the region in a cleaved slab from the photosphere to the edge of the thermal core, or where there is no thermal core, the entire slab. A comprehensive view of the gas temperature in the disk atmosphere (and the corona) can be found in Fig.~\ref{fig:T}.
Complementing this figure, Fig.~\ref{fig:Texamples} gives a detailed view of how the temperature can vary from the center of the disk to its photosphere for slabs at six diferent radii when $\dot m = 0.01$ and $M=10M_\odot$.
As it shows, and already portrayed in a different way in Fig.~\ref{fig:Tcore}, $T_g$ at the photosphere
can be anywhere from close to the temperature deep inside to an order of magnitude higher, and more often the latter than the former.

For cleaved slabs, the temperature often decreases monotonically from the photosphere to the thermal core. There is sometimes an uptick in temperature very near the core as the slab begins to thermalize. The behavior of the temperature beneath the photosphere for the cleaved slabs shown in Fig.~\ref{fig:Texamples} is opposite to the typical temperature scaling found in stellar atmospheres ($T\propto T_{\rm eff} \tau^{1/4}$). In stars, the energy generation occurs almost exclusively in the center of the star. In a black hole accretion disk, there is energy generation in the disk's core, but in these simulations there is as much or more energy generation in the disk atmosphere and in the corona outside of the photosphere (particularly in the highly spinning BH cases presented here, \citealt{Kinch_2021}). These dynamics explain the decrease of the temperature from the photosphere inwards.

Whole slabs behave differently.  They have at most moderate Thomson optical depth, anywhere from $\sim 0.1 - 10$.  Consequently, especially near the ISCO, they can have high, relatively uniform temperatures, although if the incident flux is higher on one side of the disk, then that will be reflected in the temperature profile, as in the case of $r=5\;r_g$ in Fig. \ref{fig:Texamples}. Farther out ($r=7 - 10\;r_g$), whole slabs decrease in temperature towards the midplane, but these slabs do not get cold enough to reach thermalization (see Eq. \ref{Eq:cleaving condition}). 

\begin{figure*}
\centering
\includegraphics[width=\linewidth]{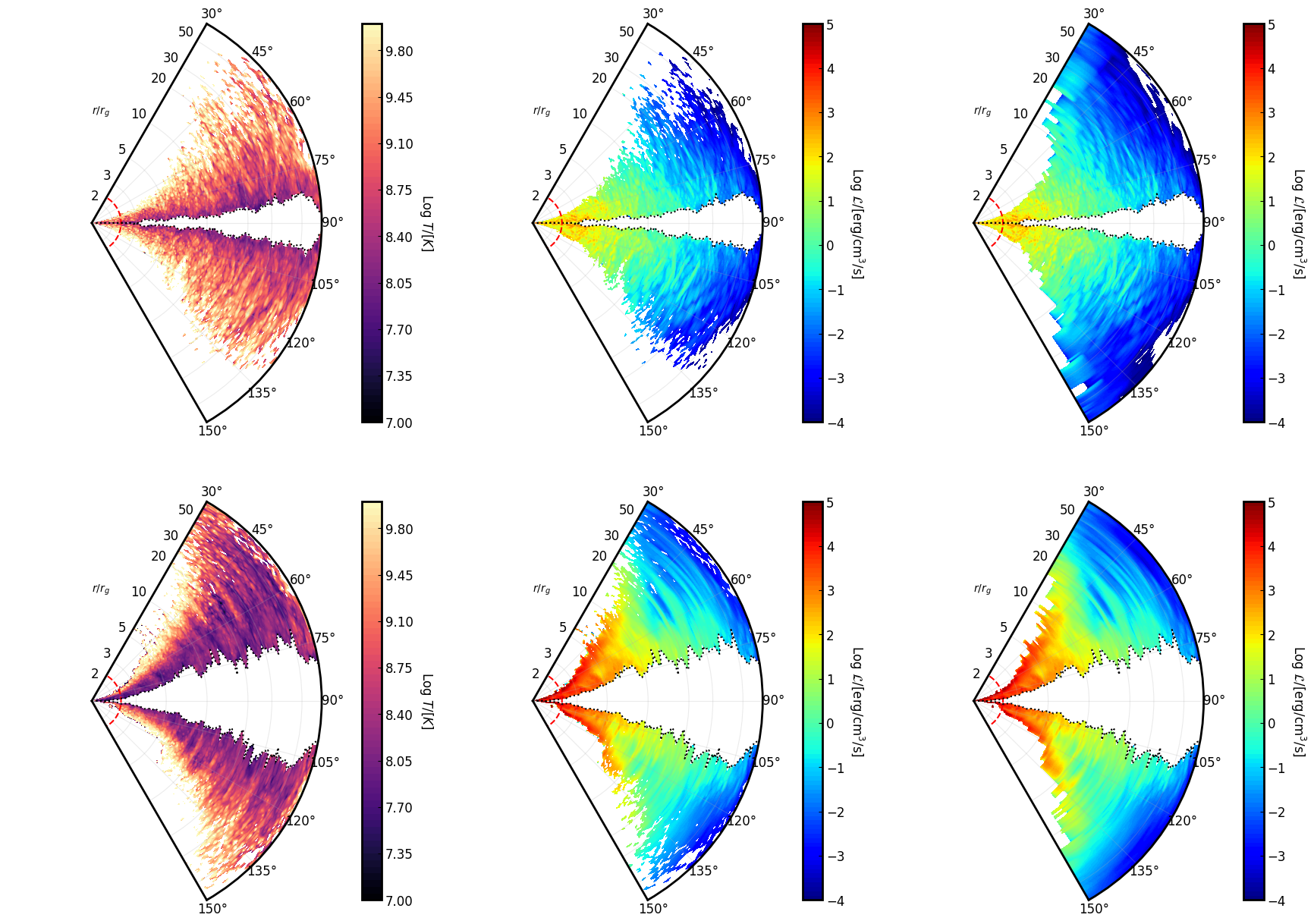}
    \caption{\texttt{Pandurata} azimuthal slices of temperature (left) and Compton power (center) and \texttt{HARM3D} dissipation rate (right) for the $M=10^8\;\msun$, $\mdot = 0.01$ model (upper) and $\mdot = 0.1$ model (lower). The red dashed line shows the ISCO while the black dotted line is the photosphere.  }
    \label{fig:maps}
\end{figure*}

\subsubsection{The corona}

We now turn from the \texttt{PTransX} solution to the region treated by \texttt{Pandurata}. The solid lines in Fig. \ref{fig:Tcore} show the mass-weighted coronal temperature. Because the density falls with distance from the disk midplane, most of the coronal mass is located near the photosphere, and the mass-weighted temperature is dominated by these cells. The temperature in the corona is hotter than the photospheric temperature, generally by a factor $\sim 3 - 10$; even this lower portion of the corona has a mean temperature $\simeq 3 \times 10^8 - 1 \times 10^9$~K.   This temperature varies with mass only slightly, with its only noticeable mass-dependence found exclusively at $r \lesssim 2 r_g$.  There is likewise hardly any trend in radius.  However, again independent of mass, the temperature away from the photosphere can rise to $\sim 10^9 - 10^{10}$~K (see Fig.~\ref{fig:maps}, left column) because the lower photon energy density cools the gas more weakly.  Because there is significant dissipative heating in these regions distant from the disk body, the luminosity-weighted mean temperature of the corona is generally $\sim 10\times$ the mass-weighted value. 

Fig. \ref{fig:maps} shows the Compton cooling power from \texttt{Pandurata} (central column) as well as the dissipation rate from \texttt{HARM3D} (right column). These two quantities are matched during the temperature iteration (Sec. \ref{sec:methods_Pandurata}), and they show excellent agreement and almost complete coverage by \texttt{Pandurata} in comparison to \texttt{HARM3D}. The only places where the coverage is not complete, due to not enough photon scatters in those cells, is at large radius and low polar angle, which are cells with very low dissipation rates.

\subsection{Energy exchange between the disk and the corona}
\label{sec:results_internal_radiation}

Having outlined the characteristic temperatures of the several subregions and explained their character, we now turn to how the two main structures, the disk and the corona, exchange energy through photons.

\subsubsection{Flux out of the photosphere}

\begin{figure*}
\centering
\includegraphics[width=\linewidth]{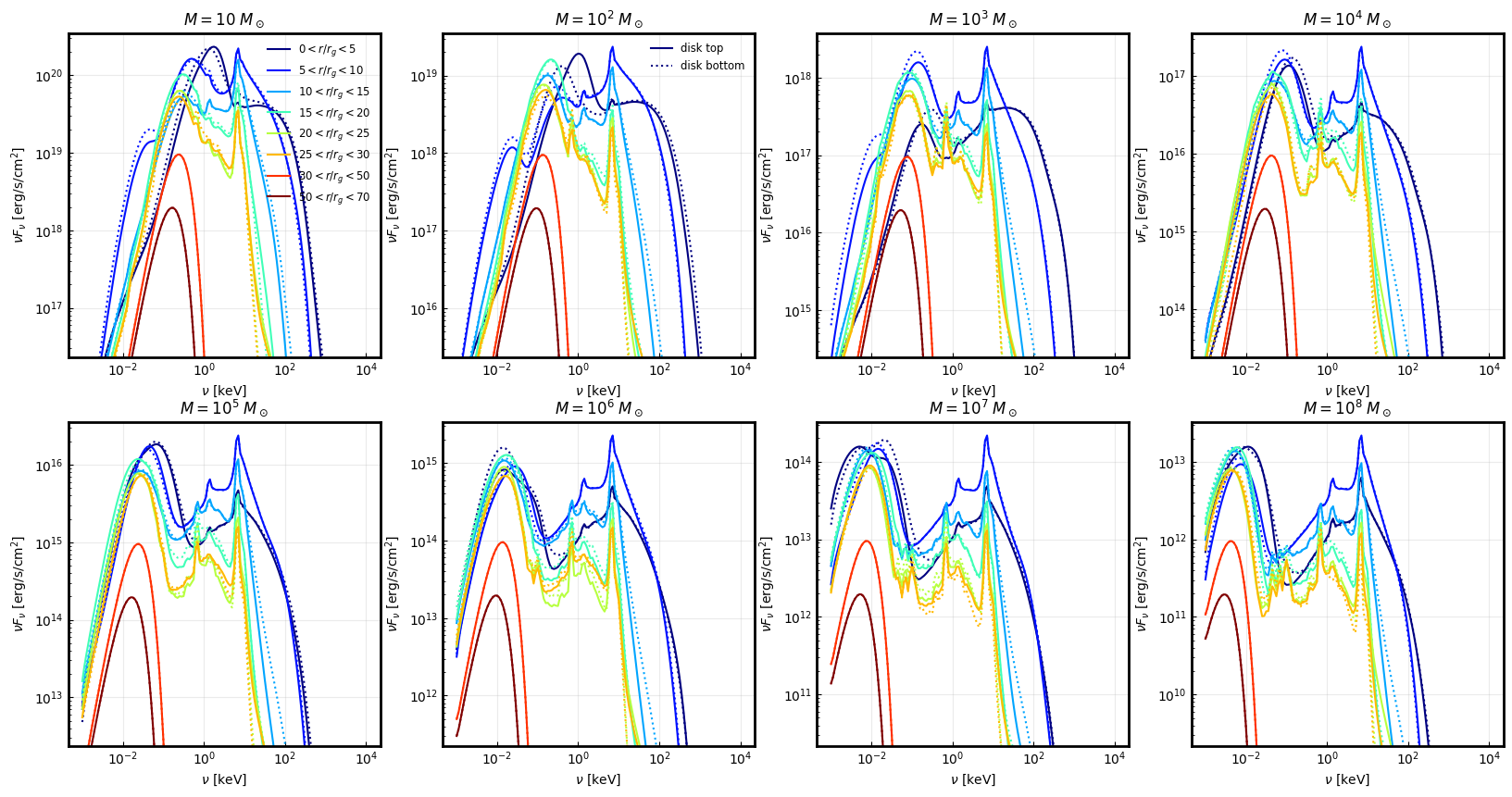}
    \caption{Spectral flux at the disk photosphere injected into the corona (computed by \texttt{PTransX}) for the $\mdot = 0.01$ case.  The spectra are grouped according to radius.  Solid and dotted curves distinguish the top and bottom surfaces.}
    \label{fig:B3}
\end{figure*}

\begin{figure*}
\centering
\includegraphics[width=\linewidth]{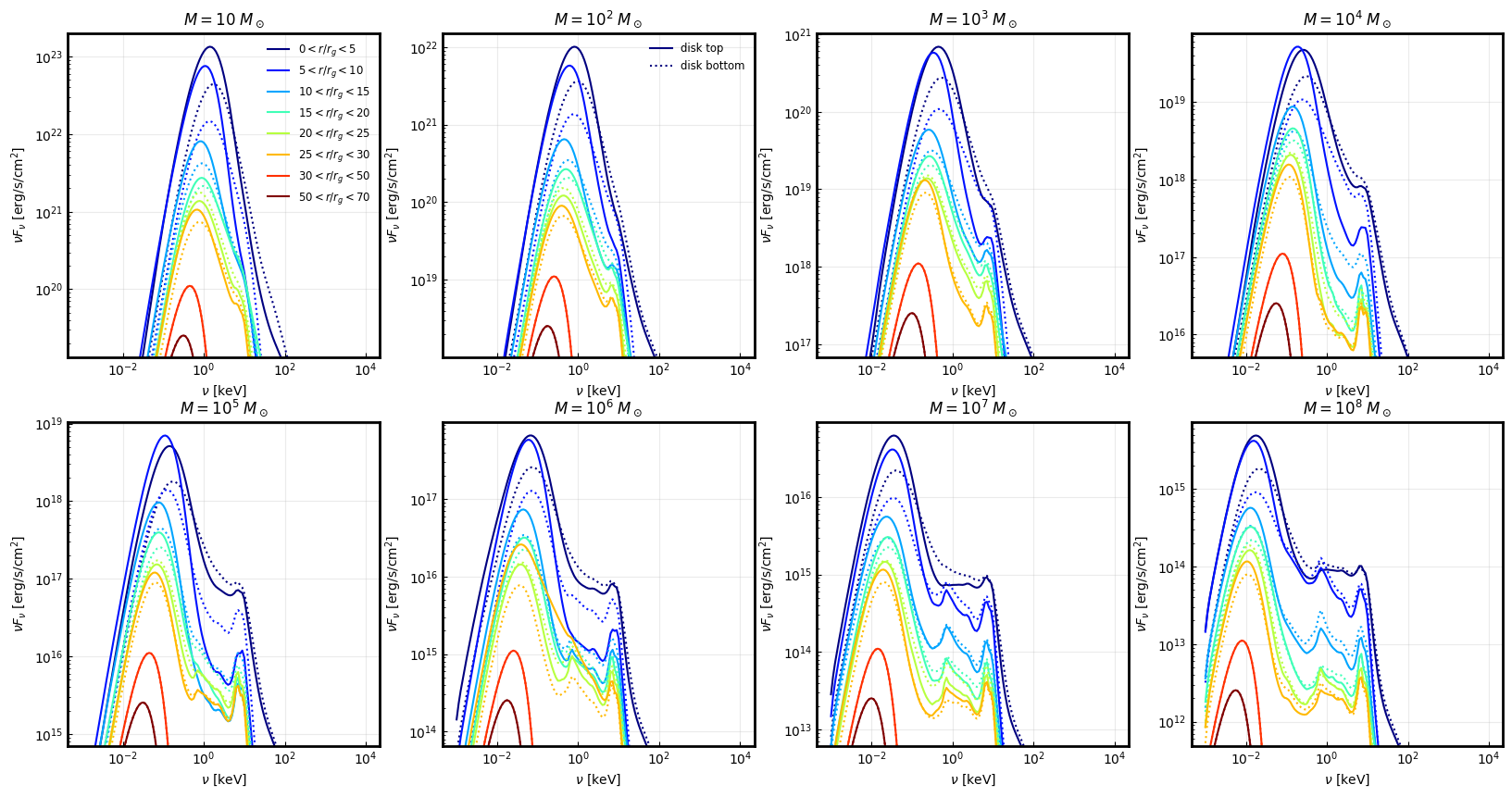}
    \caption{Spectral flux at the disk photosphere injected into the corona (computed by \texttt{PTransX}) for the $\mdot = 0.1$ case.  The spectra are grouped according to radius.  Solid and dotted curves distinguish the top and bottom surfaces.
    }
    \label{fig:B4}
\end{figure*}

Figs.~\ref{fig:B3} and \ref{fig:B4} show the spectra of the seed photon fluxes at the photosphere computed by \texttt{PTransX} for the $\mdot=0.01$ and $\mdot = 0.1$ cases, respectively.  Each panel represents a particular black hole mass; all the individual panels contain eight curves illustrating the flux averaged over different radial annuli. The vertical limits of each panel decrease by a factor of $M$. Although observed luminosity increases by a factor of $M$, the quantity plotted here is the luminosity per unit area, whose overall scaling is $\propto M^{-1}$. Both here and in our actual calculation, we define ``seed photons" as photons making their first exit from the disk into the corona.  By distinguishing them from photons that may re-enter the disk and then be reflected out, we avoid double-counting.

We will start with the $\mdot=0.01$ case.
Its most noteworthy feature is that none of the spectra produced in the \texttt{PTransX} range of radii bears any resemblance to a Planck spectrum.  In other words, no matter what the black hole mass may be, for this accretion rate at radii $\lesssim 30 r_g$, \textit{the local blackbody assumption is a very poor description of the spectrum radiated from the disk surface}.  The second-most obvious feature of Fig.~\ref{fig:B3} is how little the spectra of seed photons injected into the corona change with black hole mass.  For every mass value from $10 - 10^8 M_\odot$, each radial group's spectrum contains a thermal portion and an emission line portion of almost equal luminosity.  The central energy of the thermal portion declines with increasing mass roughly $\propto M^{-1/4}$ as expected, but the line portion, largely due to Fe~K$\alpha$, is very nearly independent of $M$.  Note, however, that the outermost two radial bins (comprising $30 < r/r_g < 70 $)
always have purely thermal disk spectra by assumption because they are outside the inflow equilibrium radius. 

We now turn to the $\mdot=0.1$ case (Fig.~\ref{fig:B4}). At this higher accretion rate, a sizable part of the flux does emerge in a quasi-blackbody spectrum, but there is an extension to higher-energy photons at $r \lesssim 15 r_g$ whose strength declines from $\sim 10\%$ of the total flux for the most massive black holes to $\sim 1\%$ of the total flux for the stellar mass black hole. Also in contrast to the $\mdot=0.01$ case, these spectra are
comparatively bereft of emission lines.  The fundamental cause of the lack of lines is that the photospheric region of the disk is farther from the midplane and therefore has lower density.  Given a flux of ionizing photons similar to the $\dot m=0.01$ case, the ionization equilibrium of the abundant elements is pushed into the regime of complete-stripping.

\begin{figure}
\centering
\includegraphics[width=\linewidth]{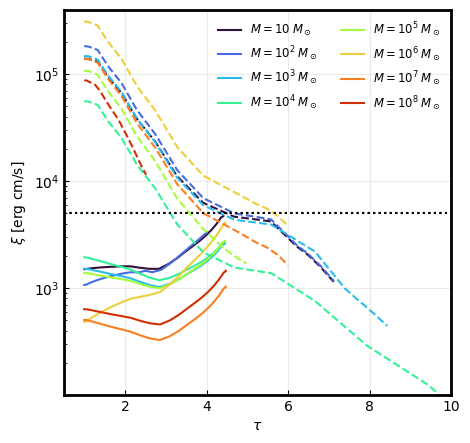}
    \caption{Ionization parameter as a function of optical depth for the $\mdot=0.01$ case (solid lines) and the $\mdot=0.1$ case (dashed lines), for a slab located at $r=15\;r_g, \phi=0$. The horizontal dotted line separates the highly ionized and fully ionized regimes.}
    \label{fig:ionization_parameter}
\end{figure}

This argument is quantified in Figure~\ref{fig:ionization_parameter}, where we show the ionization parameter $\xi$ as a function of optical depth into the disk for a single location and the full suite of black hole masses for both accretion rate cases.  We define $\xi$ as
\begin{equation}
    \xi = \frac{4 \pi \int_{7.1 \;\rm{keV}}^\infty F_{\rm{down}} d\nu}{n_e},
\end{equation}
where $F_{\rm{down}}$ is
the downward directed flux within the disk, determined from the Feautrier solution, and $n_e$ is the fluid-frame electron number density \citep{Reynolds2003PhR...377..389R}. In Fig~\ref{fig:ionization_parameter}, all $\mdot=0.01$ slabs are whole slabs, so that $\xi$ decreases towards the midplane, but then increases again once past the midplane and approaching the photosphere on the far side of the disk. In the $\mdot=0.1$ case, all slabs are upper slabs, so $\xi$ decreases monotonically with increasing $\tau$.

The horizontal dotted line in the figure is at $\xi=5000$, which approximately delineates the border between highly-ionized (below) and fully ionized (above). The figure shows that the upper atmospheres for the $\mdot=0.1$ models are fully ionized, but the entire slab in the $\mdot=0.01$ case is in the highly ionized regime, which permits creation of emission lines.
This figure only shows one example slab, but it illustrates the more general trend that for all masses, the $\mdot=0.01$ case has lower ionization degree than the $\mdot=0.1$ case, and this is true throughout the disk.
A detailed discussion of the emission line physics and its dependence on accretion rate will appear in a companion paper.

It is also worthwhile to point out two additional features of Fig.~\ref{fig:B4}. The first is the thermal peak, which decreases in energy as a function of mass (with the expected $\propto M^{-1/4}$ dependence); the second is a ``shelf" at higher energies, particularly in the inner $10\;r_g$ of higher-mass systems.
The prominence of the thermal peak demonstrates the overall importance of thermal radiation in this case, but the existence of a non-negligible higher-energy shelf demonstrates that there can also be a noticeable amount of non-thermal radiation from the disk.  This part of the disk emission comes from 
the top and bottom layers of the disk, where the gas can be strongly heated by Compton recoil or photoionization of high-energy photons entering from the corona.  This hot gas can then upscatter photons generated elsewhere within the disk.

\subsubsection{Flux onto the photosphere}

\begin{figure*}
\centering
\includegraphics[width=\linewidth]{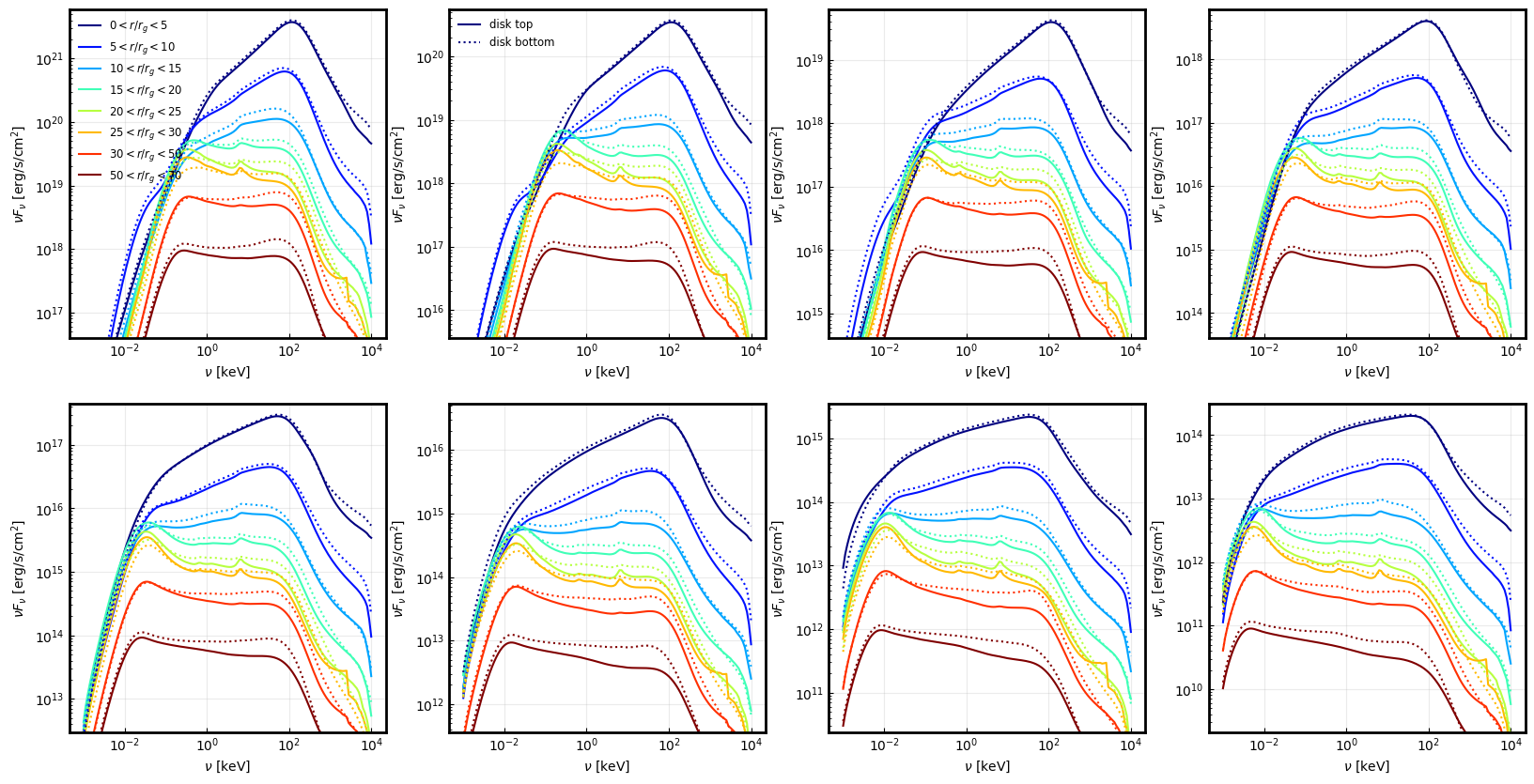}
    \caption{Spectral flux incident on the disk photosphere (computed by \texttt{Pandurata}) for the $\mdot = 0.01$ case.}
    \label{fig:B1}
\end{figure*}

\begin{figure*}
\centering
\includegraphics[width=\linewidth]{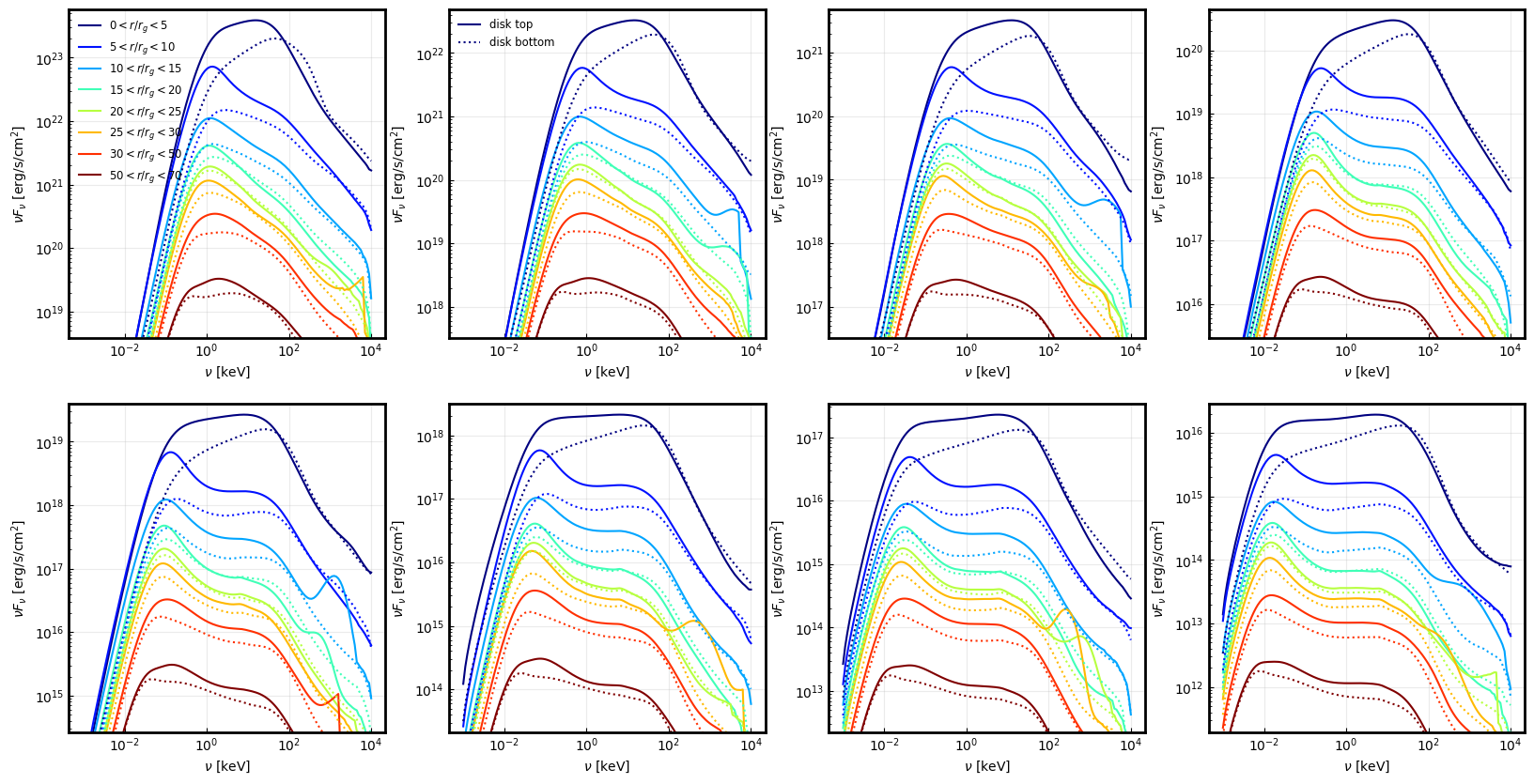}
    \caption{Spectral flux incident on the disk photosphere (computed by \texttt{Pandurata}) for the $\mdot = 0.1$ case.}
    \label{fig:B2}
\end{figure*}

Figs.~\ref{fig:B1} and \ref{fig:B2} are analogous to the figures in the previous section, but for the \texttt{Pandurata} spectrum illuminating the photosphere. Again we will start with the $\mdot=0.01$ case. The incident spectrum at each radial bin is a power law from the location of the thermal peak in Fig.~\ref{fig:B1} up to roughly 100 keV. The precise peak of the power law
declines with increasing $M$, but very slowly.
The hard X-ray power law becomes progressively ``softer" with increasing radius, showing that most hard photons are created near the black hole, as expected.

Fig. \ref{fig:B2} is the analog for the $\mdot=0.1$ case. There are three noteworthy features: a) the slope of the power law below the peak is softer than in the $\mdot=0.01$, b) the peak energy of the power law is closer to 50 keV than 100 keV, and c) there is a large disparity between the spectrum onto the upper and lower photosphere inside $10\;r_g$. The first two points are straightforward consequences of the higher mass accretion rate, which leads to less dissipation in the corona. The third point is more surprising.  It results from the combined action of two distinctions between the $\mdot = 0.01$ and $\mdot=0.1$ cases.  First, at the moment of our arbitrarily-chosen data snapshot for $\mdot = 0.1$, there is an unusually large contrast in the heating rate above and below the midplane: the heating rate in the upper hemisphere is almost twice that in the lower (a ratio of 1.78 to be exact).  Second, this heating rate contrast is sustained in the upper and lower corona spectra because the disk is optically thick almost everywhere in this higher accretion-rate simulation.   It is therefore difficult for the photon populations to mix. We believe this feature to be physical in the sense that it accurately reflects what would happen in a real accretion disk, and is consistent with the large variability seen in all energy bands of AGN observations.

\subsection{Outgoing spectra} \label{sec:results_outgoing_spectra}

As in \citep{Liu2024arXiv241201984L}, to summarize the shapes of our predicted spectra we fit simple forms to our predicted spectra over the entire range from 0.5 - 500~keV.  For lower mass cases, the form is $\nu L_\nu /L_{\rm edd} = A \nu^{2-\Gamma}$, where $\Gamma$ is the usual power-law index for the photon number spectrum.  For higher-mass cases, we include an exponential cutoff: $\nu L_\nu /L_{\rm edd} = A \nu^{2-\Gamma} \exp(-\nu/\nu_{\rm cutoff})$.
In addition, for $\mdot = 0.1$ and $M \geq 10^4 M_\odot$ (where the ``soft excess" is of greatest interest), we also fit power laws separately to the ranges 0.3~keV -- 1~keV and 1~keV -- 10~keV (we refer to these as $\Gamma_{1},\Gamma_2$, respectively).  All the best-fit parameters are reported in Table \ref{tab:fitparams}.

\begin{table}
\centering
\sisetup{scientific-notation=true, round-mode=places, round-precision=2}
\begin{tblr}{
cell{1}{1} = {r=2}{},
cell{1}{2} = {r=2}{},
cell{1}{3} = {c=6}{},
hline{1,3,11,19} = {-}{},
colspec={ccccccc}
} $\mdot$ & $\frac{M}{\msun}$& Fitting Coefficients & & & & \\
& &   A / 1000  & $\Gamma$  & $\frac{\nu_{\rm cutofff}}{\rm{keV}}$ & $\Gamma_{\rm 1}$   & $\Gamma_{\rm 2}$   & \\

$0.01$ & $10 $ & 0.13 & 1.78 & 195 & ---  & --- \\ 
$0.01$ & $10^{2} $ & 0.15 & 1.81 & 229 & ---  & --- \\ 
$0.01$ & $10^{3} $ & 0.19 & 1.83 & 217 & ---  & --- \\ 
$0.01$ & $10^{4} $ & 0.16 & 1.81 & 168 & ---  & --- \\ 
$0.01$ & $10^{5} $ & 0.23 & 1.85 & 128 & ---  & --- \\ 
$0.01$ & $10^{6} $ & 0.19 & 1.83 & 135 & ---  & --- \\ 
$0.01$ & $10^{7} $ & 0.44 & 1.95 & 126 & ---  & --- \\ 
$0.01$ & $10^{8} $ & 0.42 & 1.94 & 138 & ---  & --- \\ 
$0.1$ & $10 $ & 199 & 2.24 &  ---  & ---  & --- \\ 
$0.1$ & $10^{2} $ & 215 & 2.27 &  ---  & ---  & --- \\ 
$0.1$ & $10^{3} $ & 165 & 2.25 &  ---  & ---  & --- \\ 
$0.1$ & $10^{4} $ & 52 & 2.14 & 419& 2.46 & 2.12 \\ 
$0.1$ & $10^{5} $ & 32 & 2.09 & 233& 2.42 & 2.05 \\ 
$0.1$ & $10^{6} $ & 20 & 2.04 & 151& 2.24 & 2 \\ 
$0.1$ & $10^{7} $ & 13 & 1.98 & 94& 2.05 & 1.97 \\ 
$0.1$ & $10^{8} $ & 13 & 1.99 & 83& 1.96 & 2 \\

\end{tblr}
\caption{Fitting coefficients for spectra at different masses for $\mdot=0.01$ (upper section) and $\mdot=0.1$ (lower section). After $\mdot$ and $M$ the first three columns are for the fit to the total spectrum and the final columns ($\Gamma_1,\Gamma_2$) are for the fits demonstrating the soft excess (Fig. \ref{fig:PL_0.1}, Sec. \ref{sec:results_soft}). } \label{tab:fitparams}
\end{table}

\begin{figure}
\includegraphics[width=\linewidth]{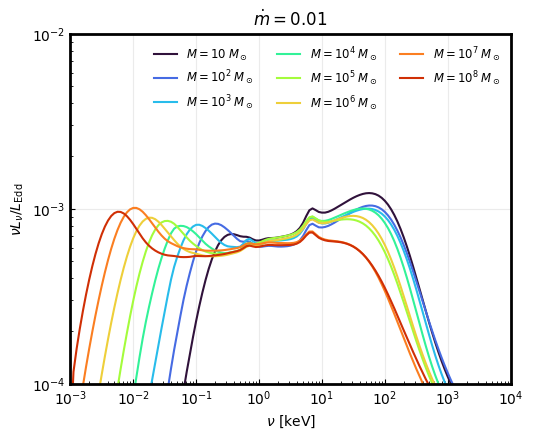}
    \caption{Emergent spectra for different masses (colors) in the $\mdot=0.01$ case as a ratio of the Eddington luminosity.}
    \label{fig:spectra_0.01}
\end{figure}

\begin{figure}
\includegraphics[width=\linewidth]{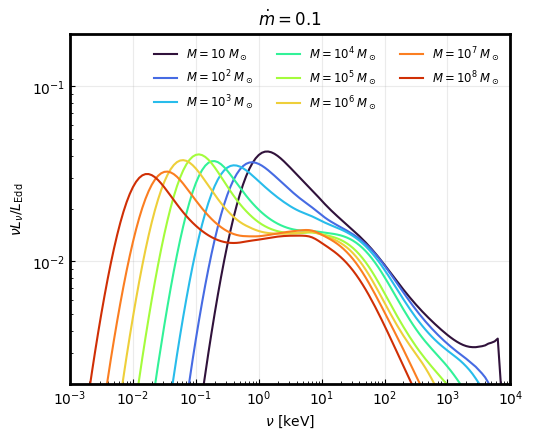}
    \caption{Same as Fig. \ref{fig:spectra_0.01} but for the $\mdot=0.1$ case.}
    \label{fig:spectra_0.1}
\end{figure}

\begin{figure}
\includegraphics[width=\linewidth]{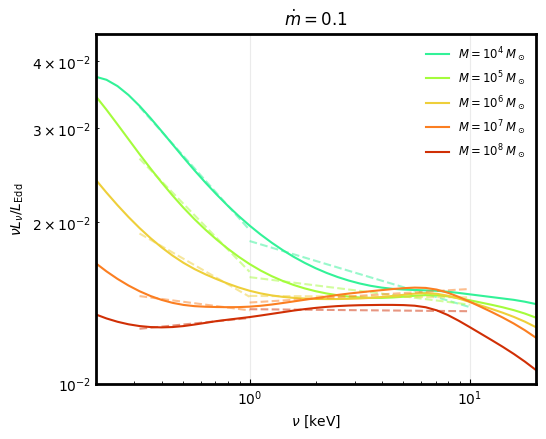}
    \caption{Same as Fig. \ref{fig:spectra_0.1} but zoomed in to show the soft X-ray excess. The dashed lines show two power fits, one from 0.3-1 keV, and the other from 1-10 keV. There is a clear change in slope from $\sim2.4$ to $2$ in these energy bands for the $M=10^{4,5}\;\msun$ models. The AGN mass models also have a change (except $M=10^8$), but it is less dramatic. The photon indices can be found in Table \ref{tab:fitparams} and a discussion of the soft excess in Sec. \ref{sec:results_soft}.}
    \label{fig:PL_0.1}
\end{figure}

\subsubsection{Continuum slope}
\label{sec:continuum_slope}

Figs.~\ref{fig:spectra_0.01} and \ref{fig:spectra_0.1} show the solid angle-integrated spectra for both accretion rate cases and all eight black hole masses.  In the $\mdot = 0.01$ case (Fig.~\ref{fig:spectra_0.01}), the power-law slopes for all black hole masses are very similar and hard, with $\Gamma \approx 1.85$ for masses below $10^7 \msun$ and $\Gamma\approx 1.95$ for the two highest mass models (Table \ref{tab:fitparams}). The slopes in the $\mdot = 0.1$ case exhibit a similarly narrow range, but somewhat softer,
with $\Gamma\approx2.15$ for $M \leq 10^5$, while the AGN masses (for the purposes of this paper, $M_{\rm AGN} = 10^{6,7,8}\;\msun$) have $\Gamma\approx 2.0 \pm 0.05$.
Thus, for lower mass models, the photon index changes with the accretion rate, but for higher mass models, in agreement with observations \citep{Ricci2017ApJS..233...17R}, the photon index does not.

The difference between
the two accretion rate cases
comes primarily from the mass-weighted coronal temperature in Fig. \ref{fig:Tcore}. This temperature in the $\mdot=0.01$ is 2-3 times higher than in the $\mdot=0.1$ case. The high energy part of the spectrum is dominated by photons that have scattered multiple times in the corona; with the temperature $2-3\times$ higher for each of these scatters, the final energy of these photons is higher by an even larger factor.

The above argument does not, however, explain why the high mass cases for $\mdot=0.1$ have flat spectra in the $\nu L_\nu$ sense, while the low mass cases are soft power-laws.
This observation can be understood by thinking of these spectra as a sum of two components: a thermal peak whose Wien tail is extended by Comptonization and a power-law with a high-energy cutoff.  As the black hole mass falls from $10^8 M_\odot$ to $10 M_\odot$, the thermal peak moves from $\sim 20$~eV to $\sim 2$~keV, while the power-law changes much less.  The result is that the power-law remains the dominant contributor down to $\sim 300$~eV for the highest black hole mass, but is overwhelmed by the Comptonized Wien tail when the black hole mass falls below $\sim 10^5 M_\odot$. When $M \lesssim 10^3 M_\odot$, the Comptonized Wien tail connects smoothly to the cutoff segment of the power-law, producing a spectrum that mimics a single soft power-law.

For $M=10\;\msun$, the higher accretion rate case ($\mdot=0.1$) has a softer spectrum, while the lower accretion rate case ($\mdot=0.01$) has a hard spectrum. 
It is sometimes postulated that the hardness of the final spectrum depends on the ratio of dissipation in the corona to the dissipation in the disk, with higher coronal dissipation leading to a harder spectrum. Here, however, the corona in both cases dominates the dissipation ($93, 91\%$ for $\mdot=0.01,0.1$ respectively). 
The coronal mass weighted temperature (Fig. \ref{fig:Tcore}) is higher in the $\mdot=0.01$ case than for $\mdot=0.1$ and
part of the reason for this difference in temperatures is that the thermal component of the seed photons has somewhat lower energy in the $\mdot=0.01$ case than for $\mdot=0.1$ (Fig. \ref{fig:Tcore}). However, 
the thermal photons are heavily diluted by non-thermal photons (Fig. \ref{fig:B3}) and the difference between the radiation temperature and the coronal temperature is several orders of magnitude, meaning that the change in electron temperature (i.e. Compton recoil) should be small ($\sim10\%$) relative to the slightly higher energy thermal photons in the $\mdot=0.1$ case.

It is not dissipation alone, however, that leads to a hard Comptonized power law. There also needs to be a sufficient density of electrons in cells with high dissipation, and it is the thermal energy of those electrons which will ultimately be imparted to the photons. If we instead compute the volume integral of the cooling rate multiplied by the density, a clear discriminant between the two cases emerges: the integral of the mass density times the luminosity density in the corona is $13\%$ of the total for $\mdot=0.01$, but only $1\%$ for $\mdot=0.1$. This diagnostic may prove useful to future GRMHD simulations.  In addition, it serves as another example of how coronae inhomogeneous in both temperature and density produce spectral shapes sensitive to factors beyond their mean temperatures and optical depths.

\subsubsection{High energy cutoff}

Observations of black holes are often fit with a high energy exponential cutoff if the observed X-ray band extends above roughly 30 keV \citep{Remillard_2006}. An inspection of our observed spectra (Figs. \ref{fig:spectra_0.01},\ref{fig:spectra_0.1}) makes it clear that instead of an exponential component at large energies, these spectra would be better fit by a broken power law.
The break occurs at $\nu_{\rm break} \sim50-80$ keV for $\mdot=0.01$ and $\sim10-50$ keV for $\mdot=0.1$,
and the slope $\Gamma$ above the break is not far from $\sim 3$. However, since most observed spectra are fitted with an exponential cutoff, that is the form we use when performing our fits (Table \ref{tab:fitparams}). We get reasonable values of the cutoff energy ($< 500$ keV) for all models in the $\mdot=0.01$ case and for $M\geq 10^4 M_\odot$ in the $\mdot=0.1$ case.

Unfortunately, more precise specification of the cutoff is hampered by two factors, one physical and one observational. The former is that \texttt{Pandurata} does not yet include pair physics. The inclusion of pair physics could significantly alter the spectrum above $\sim 100$~keV.
The observational factor is that X-ray observations are generally photon-starved at high energies, making an empirical distinction between between power laws and exponentials challenging.

\subsubsection{The low-energy thermal peak}
\label{sec:results_thermalpeak}

The angle-averaged spectra for all models show a thermal peak at low energy ($5\;\rm{eV}-1 \;\rm{keV}$). This peak is more prominent in the $\mdot=0.1$ case, and the location of the peak decreases with increasing mass. This peak is also present in the \texttt{PTransX} photospheric outgoing spectra (Figs.~\ref{fig:B3} and \ref{fig:B4}), suggesting that it originates in the disk rather than the corona. Moreover, the energy  of the peak in $\nu L_\nu$ is typically $\approx 5kT_{\rm core}$ (where we have averaged over all cleaved slabs to extract this characteristic value for $T_{\rm core}$), a signal that its photons are created within the thermal core. Whole slabs do not generally show this thermal peak, as their gas temperatures are much higher than their radiation temperatures. The only exceptions are those whole slabs whose gas temperature drops near the midplane. These slabs are on the verge of satisfying the thermalization criterion, and they produce a thermal-like flux in the colder midplane region. The thermal peak in the observed spectrum is the combination of the thermal peak in each of the slabs (binned fluxes are shown in Figs. \ref{fig:B3} and \ref{fig:B4}) with Compton broadening, area factors and relativistic shifts.

The mass dependence of the thermal peak in the outgoing spectra is clear, with $\nu_{\rm peak} \propto M^{-1/4}$ in both accretion rate cases. $\nu_{\rm peak}$ is larger in the $\mdot=0.1$ case than in the $\mdot=0.01$ case because a larger fraction of the disk is thermalized (Fig. \ref{fig:T}); with a greater amount of dissipation enclosed, $T_{\rm core}$ is larger. The amplitude of the thermal peak (in $\nu L_\nu$ space) also has a mass dependence, although the amplitude increases with mass in the $\mdot=0.01$ case and decreases with mass in the $\mdot=0.1$ case.  

As discussed in Sec.~\ref{sec:methods_PTransX}, we forced the spectrum of the radiation entering the slab from the thermal core to be a blackbody with a temperature of $T_{\rm core}$.
 If instead we used the radiation temperature (the temperature associated with the bolometric intensity) for this boundary condition,
the thermal peak in the observed spectrum would likely shift a factor $\sim 2$ upward in energy space.  This possible shift has a particular bearing on the $\mdot=0.1$ case, where the luminosity comes in large part from the thermal peak.  The location of the thermal peak also plays a critical part in the discussion of the soft X-ray excess in the next section.

\subsubsection{A soft excess for $\mdot=0.1$}
\label{sec:results_soft}

A common feature of AGN X-ray spectra is
a break in the power-law slope at $\sim 1$~keV such that $\Gamma$ changes from $\approx 2.6$ below this energy to $\approx 2$ above it. 
The steeper slope in soft X-rays is often called a ``soft excess" \citep{Chen2025arXiv250617150C}.
We find a clear soft excess for the $\mdot=0.1$ case for the $M=10^{4,5,6}\;\msun$ models.
Our models predict a similar upturn toward lower energies for AGN mass black holes, but at an energy from $\nu \sim 0.3 - 0.8$~keV, for $M=10^8 - 10^6 M_\odot $.

Independent of the black hole mass, the mechanism for this upturn can easily be identified.  It cannot be a blurring of numerous atomic lines and edges because the disk in all cases is too highly-ionized, and the disk surface spectra bear this out (Fig.~\ref{fig:B4}).  On the other hand, there are two regions in which significant Comptonization may take place, the disk atmosphere and the corona.   Their respective Compton $y$ parameters are $\sim 0.05-0.5$ and $\sim 1$. Given this comparison, Comptonization in  the corona proper emerges as the best candidate, and this identification is further  supported by the lack of significant broadening  toward  higher energies in the photospheric spectra (Fig.~\ref{fig:B4}).

\subsubsection{Angular dependence}

\begin{figure}
\includegraphics[width=\linewidth]{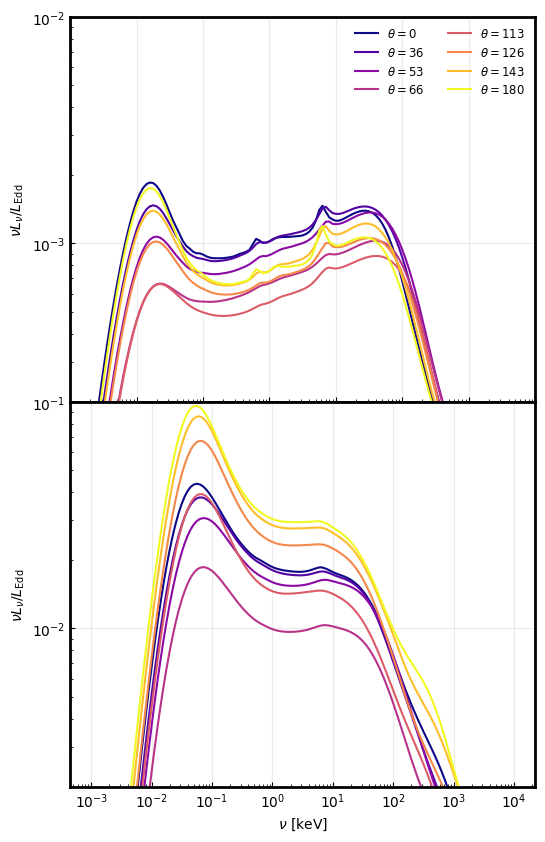}
    \caption{Spectra for different observer polar angles, as defined in the legend, where
    $\theta=0,180$ correspond to face on views of the disk.
    (Top panel) $\mdot=0.01$, $M=10^8$; (bottom panel) $\mdot = 0.1$, $M=10^6$.}
    \label{fig:spectra_angular}
\end{figure}

Fig. \ref{fig:spectra_angular} shows the polar angle dependence of the spectra in both accretion rate cases for the $M=10^6\;\msun$ model. We note that although the spectral slope and cutoff energies do not change significantly with angle, the integrated luminosity does. In particular, for the $\mdot=0.1$ case, the upper hemisphere is much more (factor of 2) luminous than the lower hemisphere, and for the $\mdot=0.01$ case, there is a smaller, but still noticeable difference. In order to better understand the origin of this difference, we reexamined the models from \citet{Liu2024arXiv241201984L}. Of the seven time snapshots (separated by 1000 M) in that paper, four have approximately equal luminosity in the upper and lower hemispheres, two have larger luminosity in the upper, and one has larger luminosity in the lower. This suggests that the behavior in Fig. \ref{fig:spectra_angular} is not atypical and there may often be half-sphere asymmetries in the light coming from black hole accretion disk systems. This up/down contrast adds another level of complication to spectral variability studies \citep{Belloni2010LNP...794...53B}. Finally we note that the line features are more prominent
in nearly face-on views.
This angle-dependence likely arises from the fact that the line features are imprinted on the disk surface and exhibit the limb-darkening of flat surfaces, whereas the continuum photons at nearby energies have mostly been scattered a number of times in the corona, whose geometry is much rounder.

\section{Discussion}  \label{sec:discussion}

\subsection{
Continuum shape; trends with $M_{\rm BH}$ and comparison to observed populations}  \label{sec:discussion_fit}

Stellar-mass accreting black holes are generally observed in X-ray binaries, which exhibit
a range of spectral states.  In the ``hard" state, most of the power emerges in a power-law whose index generally falls in the range $1.4<\Gamma<2.1$; this state is often, but not always, associated with lower luminosities \citep{Remillard_2006}.
Table \ref{tab:fitparams} shows that our stellar mass model with $\mdot=0.01$ has a photon index of $\Gamma=1.82$.  In other words, our lower luminosity $10M_\odot$ case produces a spectrum in excellent agreement with observations of the hard state. In the future we plan to compute spectra for a wider range of parameters and it will be interesting to see if a wider range of examples reproduces the spread in observed power-law slopes.

Several previous works \citep[e.g.][]{Poutanen2018A&A...614A..79P} have claimed that a coronal geometry similar to the one which comes out of \texttt{HARM3D} (that is, a clumpy, sandwich corona) should produce relatively soft spectra. The difference between our results and those of these previous studies is the high reflection fraction of the disk (Appendix B, Fig. \ref{fig:reflection}), especially the inner part that has low optical depth (Fig. \ref{fig:tau}). The high reflection fraction prevents photons from being absorbed and re-emitted as thermal photons, which would soften the spectrum. The effect of albedo on energy balance between the corona and disk and the resultant spectra has recently been studied \citep{Datta2026A&A...706A..78D}, albeit with several assumptions: a constant albedo (in energy and space), a homogeneous temperature corona (see discussion in \citealt{Liu2024arXiv241201984L}), a semi-infinite disk (not allowing for X-ray transmission through the disk), and a disk temperature which does not vary with azimuth. Despite this somewhat simplified treatment, this study also found that a higher albedo leads to a slightly harder spectrum, in line with the argument above.

The soft state is generally described by a prominent quasi-thermal peak at a few keV and a power-law spectrum at higher energies whose index is softer than that of the hard state, $\Gamma \approx 2.1 - 2.4$.  At especially high luminosities there is also a ``steep power law" state in which the entire spectrum from a few keV to $\gtrsim 100$~keV is a power-law of index $\approx 2.5$ \citep{Remillard_2006}.  Our predicted spectrum for $M=10$ and $\mdot = 0.1$ is a pure power-law with $\Gamma=2.24$; thus, it resembles the steep power law state, although it is slightly harder and somewhat lower in luminosity. \citet{Schnittman_2013} also found that the photon index increased with accretion rate from 0.01 to 0.1. However, their photon index was measured above 10 keV, higher than the typical band measured by X-ray observatories, and their sub-10 keV band was not well fit with a power law. The difference between this work and \citet{Schnittman_2013} is that we calculate the spectrum of the seed photons originating in the disk using \texttt{PTransX} instead of assuming a hardened blackbody. Especially in the inner regions of the disk, the hardened blackbody assumption does not describe the flux well (Figs. \ref{fig:B3} and \ref{fig:B4}).
To the best of of our knowledge, these predictions are the only ones
making specific statements about the spectra of luminous (i.e., $\mdot \geq 0.01$) X-ray binaries that arise directly from
global dynamical simulations and the application of radiation
transfer methods to the entire simulation domain. 

The dynamics causing a black hole X-ray binary to change spectral state remain an important open question.  This may be a topic in which the physics repertory of \texttt{HARM3D} is insufficient, a topic we will return to in the last portion of the Discussion.

The AGN mass models have spectral slopes closer to $\Gamma \approx 2$ in both accretion rate cases with the exception of the ($M=10^6\;\msun, \mdot=0.01$) model, whose spectrum is slightly harder ($\Gamma = 1.84)$.  These values fall in the middle of the observed distribution. Slightly older studies which focused on slightly more luminous AGN found that the AGN distribution peaked at $\Gamma\sim1.7-1.8$ \citep{Ricci2017ApJS..233...17R,Liu2017ApJS..232....8L,Zappacosta2018ApJ...854...33Z} while larger more recent studies have found this peak to be at $\Gamma\sim 2$ \citep{Kynoch2023MNRAS.520.2781K,Chen2025arXiv250617150C}. Thus, our results may explain the very small range of X-ray slopes exhibited by AGN of widely-differing black hole mass and luminosity.

\subsection{The soft X-ray excess}  \label{sec:discussion_soft}

AGN X-ray spectra come in many denominations, but one relatively common feature is the presence of an excess in the soft X-ray band (0.5-2 keV, known as the soft excess) relative to the underlying power law \citep{Bianchi2009A&A...495..421B,Chen2025arXiv250617150C}.
The consistent energy range of this feature has stimulated numerous suggested explanations: light-bending effects \citep{Sun1989ApJ...346...68S}; atomic absorption \citep{Gierlinski2004MNRAS.349L...7G} or emission features \citep{Crummy+2006} blurred by having a range of relativistic motions; Compton up-scattering in a warm corona \citep{Magdziarz1998MNRAS.301..179M}; or emission in the cooler outer regions of the disk \citep[$\sim100 \;r_g$][]{Jiang2025arXiv250509671J}. Recent ultraviolet observations of AGN have provided tentative support for the warm corona model \citep{Chen2025arXiv251016525C}.

As described in Sec.~\ref{sec:results_soft},
our $M=10^{4-6}\;\msun,\mdot=0.1$ models all show obvious soft excesses in the energy range 0.3-1 keV (Fig. \ref{fig:PL_0.1}).  As also remarked, the photospheric spectra for none of our $\mdot=0.1$ examples show significant atomic features, whereas the corona consistently has a Compton-$y$ parameter $\sim 1$. Consequently, we suggest that warm thermal Comptonization is the mechanism responsible for the soft excess.  This mechanism stretches the Wien tail associated with the thermal peak from lower energies to the beginning of
the ``shelf" seen in the continuum for systems with AGN masses (Sec.~\ref{sec:continuum_slope}).

The particular portion of the soft X-ray band in which this excess falls depends on $M$ through the black hole mass-dependence of the thermal peak's temperature.  When the Compton-$y$ is $\sim 1$, photon energies can be amplified by factors of several, so that a ``soft excess" is found from just above the energy of the thermal peak to energies $\sim 3 - 5\times$ higher.  Because our radiation boundary condition on the thermalization surface employs a blackbody spectrum at the minimum temperature possible at this surface, it is possible that, for $\mdot = 0.1$ and any particular mass choice, the energy range of the soft excess may be too low by a factor of a few.  Future numerical experimentation with this boundary condition can determine how large a correction may be necessary.  This same line of argument also explains why the ``soft excesses" appearing for $\mdot=0.01$ are at even lower energies ($\sim 10$~eV for the AGN masses): $T_{\rm core}$ in these cases is especially low.

It has also been suggested that the soft excess may originate from Comptonization by bulk motions rather than thermal motions in regions $\sim 50-100 r_g$ from the black hole \citep{Jiang2025arXiv250509671J}. In this study, because we perform post-processing out to $70\;r_g$,
we capture part of this region.  Even after amplification by Comptonization, we find that the portion we include contributes less than a few percent of the luminosity in the soft X-ray band.

\subsection{Comparison to models used for interpreting observations}

X-ray observations of accreting black holes are often fit with a power law component and a reflection component. This reflection component is primarily included to account for atomic emission and absorption features (which we discuss in a companion paper), but it also changes the spectral slope \citep[e.g.][]{Nandra1994MNRAS.268..405N}. It is important to note that in observationally motivated models, the reflected spectrum does not undergo Compton scattering after it is `reflected', but in reality, if the reflection occurs at an optical depth of $\tau=1$, then $\approx 2/3$ of the reflected photons will scatter at least once before leaving the corona. Our post-processing scheme naturally takes this effect into account, as `reflected' photons are injected into the corona from the disk photosphere and can then reach the observer only via \texttt{Pandurata} photon packets, which themselves undergo Compton scattering. 

Another difference between our simulations and observationally motivated models is the inhomogeneity of the emitting material. In all models in this paper, the temperature is highly inhomogeneous in both the corona and the disk body, having both large local fluctuations and global trends (Fig. \ref{fig:T}).
Simple models, however, commonly envision a single coronal temperature for a region of a single fixed Compton optical depth and a single disk surface temperature at at any given radius. In \citet{Liu2024arXiv241201984L}, we showed that a single temperature corona could not be tuned to reproduce the spectrum of an inhomogeneous temperature corona found by \texttt{Pandurata/PTransX} (Fig. 6 of \citealt{Liu2024arXiv241201984L}). Regarding the treatment of the disk as having a single temperature, Figs. \ref{fig:B3}, \ref{fig:B4} clearly show that the photospheric flux has different peaks at different locations on the disk surface. Although the emergent spectra (Figs. \ref{fig:spectra_0.01}, \ref{fig:spectra_0.1}) all have a single thermal peak,
this temperature arises out of a complex spectral averaging process that is not a simple weighted average of disk surface temperatures.

This latter point is more general.  Because the emergent spectrum is such a complex and nonlinear function of simple variables like optical depth, temperature, etc., it {\it cannot} be found by solving the transfer problem for a single region whose parameters are averages over the entire system.

\subsection{Model limitations and future directions}

In this paper, we have explored the effects of two parameters, mass and mass accretion rate, on black hole spectra. Many more interesting parameters exist. The black hole spin is a frequently studied parameter as it affects dynamics close to the horizon, for example the strength of jets. The target entropy in the GRMHD simulation controls the
pressure throughout the disk body; changing 
the cooling rate in any way would have repercussions in disk structure, both radial and vertical (e.g., the scale height), that could lead to changes in the seed photon spectrum, and therefore both the continuum and lines.
There is also the effect of initial magnetic field geometry and net flux; fields whose basic topology is toroidal lead to different disk structures than those that are poloidal \citep{BHK2008};
magnetically-dominated disks have their own set of distinct structural properties \citep[e.g.][]{Dhang2025ApJ...980..203D}.
All of these parameters should be investigated more fully by future work, but there are also improvements to be realized in both the underlying simulations and the post-processing. 

Part of what we will learn from this program is which elements of the physics repertory in a simulation code are necessary for \texttt{PTransX/Pandurata} to produce a reliable answer for which observable properties. In this paper, we have demonstrated how postprocessing of \texttt{HARM3D} snapshots reproduces some, but not all, observed spectral features.

Looking toward future work, we highlight physical mechanisms whose inclusion in GRMHD simulations may, in our view, have the greatest potential to alter the spectral features determined by the postprocessing. The first of these is multi-angle radiation transfer \citep{White2023ApJ...949..103W,Zhang2025arXiv250602289Z}, which allows radiation forces to be accounted for dynamically and, even when radiation forces are unimportant, yields a physically credible cooling rate (see above remarks about cooling).  A more accurate gas pressure, plus allowance for radiation pressure, will give more reliable results for many things, including the vertical density profile of the disk, often condensed into the parameter $H/R$. A subsequent improvement to GRMHD codes will be spectral radiation transfer, as the radiation around black hole accretion disks is highly non thermal (as we have shown).  However, for the foreseeable future, the number of photon-energy groups that can be treated will be small enough that the result will be a rather rough approximation to the continuum shape. Next, the inclusion of electron-positron pairs could significantly alter both the Compton opacity and the thermodynamics of the corona. Heat conduction from the corona to the disk has been suggested as a possible influence on disk thermodynamics, although this is usually considered a secondary effect \citep{Maciolek-Niedzwiecki1997ApJ...483..111M,Bambic2024MNRAS.530.1812B}.

Finally, we list some future improvements which could be made to the post-processing simulations. One improvement is a better way to determine the location of the thermalization surface and the spectrum of the net flux emerging from the thermalized region in a fashion that accounts for the typically small ratio of flux to intensity at that boundary. Another is the inclusion of electron-positron pairs and/or non-thermal electron distributions in low density regions of the corona. And lastly we hope to realize improvements in resolution (or equivalently, computing time/speed) so as to match the resolution achieved by the GRMHD simulations.

\section{Conclusions} 
\label{sec:conclusion}

In this paper, we have post-processed data from GRMHD simulations of rapidly spinning black holes accreting material in the radiatively efficient regime. We have done this for a regular grid spanning seven orders of magnitude in black hole mass.
The first accomplishment of this paper is that we have:

\begin{itemize}

\item Demonstrated that \texttt{PTransX/Pandurata} can achieve a global solution for temperature, ionization state, and the multi-frequency radiation transfer problem for black hole masses from $10M_\odot$ to $10^8 M_\odot$ using as input a 3D snapshot of density, velocity, and cooling rate from a GRMHD simulation.

\end{itemize}
Our previous work had demonstrated that achieving global thermal balance in black hole accretion systems was possible at $M=10^{} \;\msun$, but in order to extend this solution to higher masses, large sections of our method needed to be reexamined and modified. Although there are many changes that have been made, we highlight two important ones: increased \texttt{PTransX} coverage of the inner part of the accretion disk, and a restructuring of the thermalization criterion in order to reflect the underlying physics and energy conservation in each slab. The increased \texttt{PTransX} coverage gives a much improved seed photon spectrum near the inner edge of the disk, a quantity crucial to the determination of the final spectral slope. The thermalization criterion is important in determining the location and strength of the thermal peak that appears in the radiated spectrum. 

In addition, we have:

\begin{itemize}

\item Confirmed for a wide range of black hole masses \citet{Liu2024arXiv241201984L}'s demonstration that Comptonizing coronae comprise a wide range of temperatures.

\end{itemize}
Fig. \ref{fig:T} shows this fact clearly for each of the models in this paper. As shown in \citet{Liu2024arXiv241201984L}, assuming a single temperature corona cannot reproduce the final spectrum for the \texttt{HARM3D} snapshots studied in that work, yet this assumption underpins many observational studies of the coronal geometry.

\begin{itemize}

\item Shown that at an accretion rate as high as $\mdot = 0.01$ the spectrum emitted from the disk's photosphere is very different from a blackbody, although at higher accretion rates the quality of this approximation improves.

\end{itemize}

\begin{itemize}
\item Found that well-known radiation mechanisms applied to GRMHD simulation data produced with \texttt{HARM3D} for a rapidly spinning black hole can replicate the  spectrum of both the hard and steep power-law states of $10M_\odot$ black holes, with the steep power-law appearing at higher accretion rate than the hard state.
\end{itemize}

The root cause of the different spectral states of stellar-mass black holes
is one of the biggest open questions in the study of accreting black holes. We have shown that
increasing the accretion rate a factor of 10 can account for a transition from the hard state to the steep power-law state.

\begin{itemize}

\item Found, using the same well-known radiation mechanisms, that GRMHD simulation data produced with \texttt{HARM3D} can replicate the X-ray spectrum of AGN for $10^6 - 10^8M_\odot$ black holes over a range of accretion rates.

\end{itemize}
Our solutions are also consistent with the fact that hard X-ray observations ($\nu>10$ keV) of AGN tend to find softer photon indices than observations spanning the entire X-ray band \citep{Ricci2017ApJS..233...17R}. This conclusion applies to both of our accretion rate cases. 

\begin{itemize}

\item Shown that the disk's thermal peak can be broadened into a ``soft X-ray excess" by warm Comptonization, but not by blending atomic features.  This feature is clearest at higher accretion rates, and its onset shifts in energy as a function of black hole mass.
\end{itemize}

The origin of the soft excess has been a persistent mystery in the study of accreting black holes. For the first time, we have shown how physical accretion flows along with a careful treatment of radiation processes can produce it.
Although the results presented here indicate that for black hole masses in the AGN range the soft excess begins at photon energies a factor of several too low, there is reason to think that this discrepancy may be an artifact of a boundary condition
(Sec. \ref{sec:discussion_soft}).

\begin{itemize}

\item Finally, and most importantly, we have shown that the generic assumption of MHD-driven accretion leads directly to radiated spectra strongly resembling those observed. Although the GRMHD simulations have free parameters, such as the disk scale height and magnetic field geometry, the post-processing was performed {\it without a single free parameter}. 
\end{itemize}

In other words, we can start from first principles in terms of dynamics, introduce only the atomic data needed for the various opacities, ionization equilibria, etc., and arrive at a first approximation to real spectra.  This success in connecting GRMHD simulations to observable spectral features suggests that the underlying conceptual model of accretion taking place through the action of fluctuating, but correlated, magnetic stresses says much that is correct about how accretion flows onto black holes generate light.

It is, however, incomplete.  There are important elements of black hole accretion phenomenology that the present work is unable to explain or leaves unexplored.  We have not, for example, shown how the soft state of stellar-mass X-ray binaries is generated.  We have also not been able to extend our predicted spectra for AGN down to the optical/UV range. Nor does the soft excess appear in the correct band for $M\geq10^7\;\msun$.  It also remains to be seen whether all these results obtain for black hole spin parameters other than 0.9.  We regard these gaps as pointers toward improvements in the implementation of this model in both the underlying dynamical simulations and the post-processing.

\begin{center}  \label{Sec:ack}
    \textbf{Acknowledgments}
\end{center}

We thank Scott Noble, the author of \texttt{HARM3D}, for use of his code and for much help in running it.

This work was partially supported by NSF Grant AST-2009260 and NASA TCAN grant 80NSSC24K0100.

\bibliography{references}{}

@ARTICLE{Datta2026A&A...706A..78D,
       author = {{Datta}, Sudeb Ranjan and {Bursa}, Michal and {Dovciak}, Michal and {Zhang}, Wenda},
        title = "{Interaction between disk and extended corona in a general relativistic framework: I. Static slab corona in global energy balance with the underlying disk}",
      journal = {\aap},
         year = 2026,
        month = feb,
       volume = {706},
          eid = {A78},
        pages = {A78},
          doi = {10.1051/0004-6361/202554464},
archivePrefix = {arXiv},
       eprint = {2509.10177},
 primaryClass = {astro-ph.HE},
       adsurl = {https://ui.adsabs.harvard.edu/abs/2026A&A...706A..78D}
}

@ARTICLE{Nandra1994MNRAS.268..405N,
       author = {{Nandra}, K. and {Pounds}, K.~A.},
        title = "{GINGA observations of the X-ray spectra of Seyfert galaxies.}",
      journal = {\mnras},
         year = 1994,
        month = may,
       volume = {268},
        pages = {405-429},
          doi = {10.1093/mnras/268.2.405},
       adsurl = {https://ui.adsabs.harvard.edu/abs/1994MNRAS.268..405N}
}

@ARTICLE{Poutanen2018A&A...614A..79P,
       author = {{Poutanen}, Juri and {Veledina}, Alexandra and {Zdziarski}, Andrzej A.},
        title = "{Doughnut strikes sandwich: the geometry of hot medium in accreting black hole X-ray binaries}",
      journal = {\aap},
         year = 2018,
        month = jun,
       volume = {614},
          eid = {A79},
        pages = {A79},
          doi = {10.1051/0004-6361/201732345},
archivePrefix = {arXiv},
       eprint = {1711.08509},
 primaryClass = {astro-ph.HE},
       adsurl = {https://ui.adsabs.harvard.edu/abs/2018A&A...614A..79P}
}

@ARTICLE{Maciolek-Niedzwiecki1997ApJ...483..111M,
       author = {{Macio{\L}ek-Nied{\'z}wiecki}, Andrzej and {Krolik}, Julian H. and {Zdziarski}, Andrzej A.},
        title = "{Thermal Conduction in Accretion Disk Coronae}",
      journal = {\apj},
         year = 1997,
        month = jul,
       volume = {483},
       number = {1},
        pages = {111-120},
          doi = {10.1086/304225},
archivePrefix = {arXiv},
       eprint = {astro-ph/9702068},
 primaryClass = {astro-ph},
       adsurl = {https://ui.adsabs.harvard.edu/abs/1997ApJ...483..111M}
}

@ARTICLE{Bambic2024MNRAS.530.1812B,
       author = {{Bambic}, Christopher J. and {Quataert}, Eliot and {Kunz}, Matthew W. and {Jiang}, Yan-Fei},
        title = "{Local models of two-temperature accretion disc coronae - II. Ion thermal conduction and the absence of disc evaporation}",
      journal = {\mnras},
         year = 2024,
        month = may,
       volume = {530},
       number = {2},
        pages = {1812-1828},
          doi = {10.1093/mnras/stae696},
archivePrefix = {arXiv},
       eprint = {2401.05482},
 primaryClass = {astro-ph.HE},
       adsurl = {https://ui.adsabs.harvard.edu/abs/2024MNRAS.530.1812B}
}

@ARTICLE{Huang2025arXiv251212728H,
       author = {{Huang}, Yimin and {Liu}, Honghui and {Bambi}, Cosimo and {Ingram}, Adam and {Jiang}, Jiachen and {Young}, Andrew and {Zhang}, Zuobin},
        title = "{DAO: A New and Public Non-Relativistic Reflection Model}",
      journal = {arXiv e-prints},
         year = 2025,
        month = dec,
          eid = {arXiv:2512.12728},
        pages = {arXiv:2512.12728},
archivePrefix = {arXiv},
       eprint = {2512.12728},
 primaryClass = {astro-ph.HE},
       adsurl = {https://ui.adsabs.harvard.edu/abs/2025arXiv251212728H}
}

@ARTICLE{Chen2025arXiv251016525C,
       author = {{Chen}, Shi-Jiang and {Wang}, Jun-Xian and {Kang}, Jia-Lai and {Kang}, Wen-Yong and {Sou}, Hao and {Liu}, Teng and {Cai}, Zhen-Yi and {Su}, Zhen-Bo},
        title = "{A UV to X-Ray View of Soft Excess in Type 1 Active Galactic Nuclei. II. Broadband Correlations}",
      journal = {arXiv e-prints},
         year = 2025,
        month = oct,
          eid = {arXiv:2510.16525},
        pages = {arXiv:2510.16525},
archivePrefix = {arXiv},
       eprint = {2510.16525},
 primaryClass = {astro-ph.HE},
       adsurl = {https://ui.adsabs.harvard.edu/abs/2025arXiv251016525C}
}

@ARTICLE{Sun1989ApJ...346...68S,
       author = {{Sun}, Wei-Hsin and {Malkan}, Matthew A.},
        title = "{Fitting Improved Accretion Disk Models to the Multiwavelength Continua of Quasars and Active Galactic Nuclei}",
      journal = {\apj},
         year = 1989,
        month = nov,
       volume = {346},
        pages = {68},
          doi = {10.1086/167986},
       adsurl = {https://ui.adsabs.harvard.edu/abs/1989ApJ...346...68S}
}

@INCOLLECTION{Belloni2010LNP...794...53B,
       author = {{Belloni}, T.~M.},
        title = "{States and Transitions in Black Hole Binaries}",
    booktitle = {Lecture Notes in Physics, Berlin Springer Verlag},
         year = 2010,
       editor = {{Belloni}, Tomaso},
       volume = {794},
        pages = {53},
          doi = {10.1007/978-3-540-76937-8_3},
       adsurl = {https://ui.adsabs.harvard.edu/abs/2010LNP...794...53B}
}

@ARTICLE{RanjanDatta2025arXiv250910177R,
       author = {{Ranjan Datta}, Sudeb and {Bursa}, Michal and {Dovciak}, Michal and {Zhang}, Wenda},
        title = "{Interaction between disk and extended corona in general relativistic framework. I. Static slab corona in global energy balance with the underlying disk}",
      journal = {arXiv e-prints},
         year = 2025,
        month = sep,
          eid = {arXiv:2509.10177},
        pages = {arXiv:2509.10177},
archivePrefix = {arXiv},
       eprint = {2509.10177},
 primaryClass = {astro-ph.HE},
       adsurl = {https://ui.adsabs.harvard.edu/abs/2025arXiv250910177R}
}

@ARTICLE{Shashank2025arXiv250702583S,
       author = {{Shashank}, Swarnim and {Abdikamalov}, Askar B. and {Liu}, Honghui and {Nosirov}, Abdurakhmon and {Bambi}, Cosimo and {Dihingia}, Indu K. and {Mizuno}, Yosuke},
        title = "{Measuring black hole spins with X-ray reflection spectroscopy: A GRMHD outlook}",
      journal = {arXiv e-prints},
         year = 2025,
        month = jul,
          eid = {arXiv:2507.02583},
        pages = {arXiv:2507.02583},
          doi = {10.48550/arXiv.2507.02583},
archivePrefix = {arXiv},
       eprint = {2507.02583},
 primaryClass = {astro-ph.HE},
       adsurl = {https://ui.adsabs.harvard.edu/abs/2025arXiv250702583S}
}

@ARTICLE{Liu2017ApJS..232....8L,
       author = {{Liu}, Teng and {Tozzi}, Paolo and {Wang}, Jun-Xian and {Brandt}, William N. and {Vignali}, Cristian and {Xue}, Yongquan and {Schneider}, Donald P. and {Comastri}, Andrea and {Yang}, Guang and {Bauer}, Franz E. and {Paolillo}, Maurizio and {Luo}, Bin and {Gilli}, Roberto and {Wang}, Q. Daniel and {Giavalisco}, Mauro and {Ji}, Zhiyuan and {Alexander}, David M. and {Mainieri}, Vincenzo and {Shemmer}, Ohad and {Koekemoer}, Anton and {Risaliti}, Guido},
        title = "{X-Ray Spectral Analyses of AGNs from the 7Ms Chandra Deep Field-South Survey: The Distribution, Variability, and Evolutions of AGN Obscuration}",
      journal = {\apjs},
         year = 2017,
        month = sep,
       volume = {232},
       number = {1},
          eid = {8},
        pages = {8},
          doi = {10.3847/1538-4365/aa7847},
archivePrefix = {arXiv},
       eprint = {1703.00657},
 primaryClass = {astro-ph.GA},
       adsurl = {https://ui.adsabs.harvard.edu/abs/2017ApJS..232....8L}
}

@ARTICLE{Zappacosta2018ApJ...854...33Z,
       author = {{Zappacosta}, L. and {Comastri}, A. and {Civano}, F. and {Puccetti}, S. and {Fiore}, F. and {Aird}, J. and {Del Moro}, A. and {Lansbury}, G.~B. and {Lanzuisi}, G. and {Goulding}, A. and {Mullaney}, J.~R. and {Stern}, D. and {Ajello}, M. and {Alexander}, D.~M. and {Ballantyne}, D.~R. and {Bauer}, F.~E. and {Brandt}, W.~N. and {Chen}, C. -T.~J. and {Farrah}, D. and {Harrison}, F.~A. and {Gandhi}, P. and {Lanz}, L. and {Masini}, A. and {Marchesi}, S. and {Ricci}, C. and {Treister}, E.},
        title = "{The NuSTAR  Extragalactic Surveys: X-Ray Spectroscopic Analysis of the Bright Hard-band Selected Sample}",
      journal = {\apj},
         year = 2018,
        month = feb,
       volume = {854},
       number = {1},
          eid = {33},
        pages = {33},
          doi = {10.3847/1538-4357/aaa550},
archivePrefix = {arXiv},
       eprint = {1801.04280},
 primaryClass = {astro-ph.HE},
       adsurl = {https://ui.adsabs.harvard.edu/abs/2018ApJ...854...33Z}
}

@ARTICLE{Kynoch2023MNRAS.520.2781K,
       author = {{Kynoch}, Daniel and {Mitchell}, Jake A.~J. and {Ward}, Martin J. and {Done}, Chris and {Lusso}, Elisabeta and {Landt}, Hermine},
        title = "{The SOUX AGN sample: SDSS-XMM-Newton optical, ultraviolet, and X-ray selected active galactic nuclei spanning a wide range of parameter space - sample definition}",
      journal = {\mnras},
         year = 2023,
        month = apr,
       volume = {520},
       number = {2},
        pages = {2781-2805},
          doi = {10.1093/mnras/stad221},
archivePrefix = {arXiv},
       eprint = {2301.07724},
 primaryClass = {astro-ph.GA},
       adsurl = {https://ui.adsabs.harvard.edu/abs/2023MNRAS.520.2781K}
}

@ARTICLE{Zhang2025arXiv250602289Z,
       author = {{Zhang}, Lizhong and {Stone}, James M. and {Mullen}, Patrick D. and {Davis}, Shane W. and {Jiang}, Yan-Fei and {White}, Christopher J.},
        title = "{Radiation GRMHD Models of Accretion onto Stellar-Mass Black Holes: I. Survey of Eddington Ratios}",
      journal = {arXiv e-prints},
         year = 2025,
        month = jun,
          eid = {arXiv:2506.02289},
        pages = {arXiv:2506.02289},
          doi = {10.48550/arXiv.2506.02289},
archivePrefix = {arXiv},
       eprint = {2506.02289},
 primaryClass = {astro-ph.HE},
       adsurl = {https://ui.adsabs.harvard.edu/abs/2025arXiv250602289Z}
}

@ARTICLE{Liska2022ApJ...935L...1L,
       author = {{Liska}, M.~T.~P. and {Musoke}, G. and {Tchekhovskoy}, A. and {Porth}, O. and {Beloborodov}, A.~M.},
        title = "{Formation of Magnetically Truncated Accretion Disks in 3D Radiation-transport Two-temperature GRMHD Simulations}",
      journal = {\apjl},
         year = 2022,
        month = aug,
       volume = {935},
       number = {1},
          eid = {L1},
        pages = {L1},
          doi = {10.3847/2041-8213/ac84db},
archivePrefix = {arXiv},
       eprint = {2201.03526},
 primaryClass = {astro-ph.HE},
       adsurl = {https://ui.adsabs.harvard.edu/abs/2022ApJ...935L...1L}
}

@ARTICLE{NaetheMotta2025arXiv250508855N,
       author = {{Naethe Motta}, Pedro and {Jacquemin-Ide}, Jonatan and {Nemmen}, Rodrigo and {Liska}, Matthew T.~P. and {Tchekhovskoy}, Alexander},
        title = "{Black hole spectral states revealed in GRMHD simulations with texture memory accelerated cooling}",
      journal = {arXiv e-prints},
         year = 2025,
        month = may,
          eid = {arXiv:2505.08855},
        pages = {arXiv:2505.08855},
          doi = {10.48550/arXiv.2505.08855},
archivePrefix = {arXiv},
       eprint = {2505.08855},
 primaryClass = {astro-ph.HE},
       adsurl = {https://ui.adsabs.harvard.edu/abs/2025arXiv250508855N}
}

@ARTICLE{McConnell2013ApJ...764..184M,
       author = {{McConnell}, Nicholas J. and {Ma}, Chung-Pei},
        title = "{Revisiting the Scaling Relations of Black Hole Masses and Host Galaxy Properties}",
      journal = {\apj},
         year = 2013,
        month = feb,
       volume = {764},
       number = {2},
          eid = {184},
        pages = {184},
          doi = {10.1088/0004-637X/764/2/184},
archivePrefix = {arXiv},
       eprint = {1211.2816},
 primaryClass = {astro-ph.CO},
       adsurl = {https://ui.adsabs.harvard.edu/abs/2013ApJ...764..184M}
}

@ARTICLE{Dhang2025ApJ...980..203D,
       author = {{Dhang}, Prasun and {Dexter}, Jason and {Begelman}, Mitchell C.},
        title = "{Energy Extraction from a Black Hole by a Strongly Magnetized Thin Accretion Disk}",
      journal = {\apj},
         year = 2025,
        month = feb,
       volume = {980},
       number = {2},
          eid = {203},
        pages = {203},
          doi = {10.3847/1538-4357/ada76e},
archivePrefix = {arXiv},
       eprint = {2411.02515},
 primaryClass = {astro-ph.HE},
       adsurl = {https://ui.adsabs.harvard.edu/abs/2025ApJ...980..203D}
}

@ARTICLE{Weisskopf2002PASP..114....1W,
       author = {{Weisskopf}, M.~C. and {Brinkman}, B. and {Canizares}, C. and {Garmire}, G. and {Murray}, S. and {Van Speybroeck}, L.~P.},
        title = "{An Overview of the Performance and Scientific Results from the Chandra X-Ray Observatory}",
      journal = {\pasp},
         year = 2002,
        month = jan,
       volume = {114},
       number = {791},
        pages = {1-24},
          doi = {10.1086/338108},
archivePrefix = {arXiv},
       eprint = {astro-ph/0110308},
 primaryClass = {astro-ph},
       adsurl = {https://ui.adsabs.harvard.edu/abs/2002PASP..114....1W}
}

@ARTICLE{Chen2025arXiv250617150C,
       author = {{Chen}, Shi-Jiang and {Buchner}, Johannes and {Liu}, Teng and {Hagen}, Scott and {Waddell}, Sophia G.~H. and {Nandra}, Kirpal and {Salvato}, Mara and {Igo}, Zsofi and {Aydar}, Catarina and {Merloni}, Andrea and {Ni}, Qingling and {Kang}, Jia-Lai and {Cai}, Zhen-Yi and {Wang}, Jun-Xian and {Li}, Ruancun and {Ramos-Ceja}, Miriam E. and {Sanders}, Jeremy and {Georgakakis}, Antonis and {Zhang}, Yi},
        title = "{The Average Soft X-ray Spectra of eROSITA Active Galactic Nuclei}",
      journal = {arXiv e-prints},
         year = 2025,
        month = jun,
          eid = {arXiv:2506.17150},
        pages = {arXiv:2506.17150},
          doi = {10.48550/arXiv.2506.17150},
archivePrefix = {arXiv},
       eprint = {2506.17150},
 primaryClass = {astro-ph.HE},
       adsurl = {https://ui.adsabs.harvard.edu/abs/2025arXiv250617150C}
}

@ARTICLE{Ricci2017ApJS..233...17R,
       author = {{Ricci}, C. and {Trakhtenbrot}, B. and {Koss}, M.~J. and {Ueda}, Y. and {Delvecchio}, I. and {Treister}, E. and {Schawinski}, K. and {Paltani}, S. and {Oh}, K. and {Lamperti}, I. and {Berney}, S. and {Gandhi}, P. and {Ichikawa}, K. and {Bauer}, F.~E. and {Ho}, L.~C. and {Asmus}, D. and {Beckmann}, V. and {Soldi}, S. and {Balokovi{\'c}}, M. and {Gehrels}, N. and {Markwardt}, C.~B.},
        title = "{BAT AGN Spectroscopic Survey. V. X-Ray Properties of the Swift/BAT 70-month AGN Catalog}",
      journal = {\apjs},
         year = 2017,
        month = dec,
       volume = {233},
       number = {2},
          eid = {17},
        pages = {17},
          doi = {10.3847/1538-4365/aa96ad},
archivePrefix = {arXiv},
       eprint = {1709.03989},
 primaryClass = {astro-ph.HE},
       adsurl = {https://ui.adsabs.harvard.edu/abs/2017ApJS..233...17R}
}

@ARTICLE{Gehrels2004ApJ...611.1005G,
       author = {{Gehrels}, N. and {Chincarini}, G. and {Giommi}, P. and {Mason}, K.~O. and {Nousek}, J.~A. and {Wells}, A.~A. and {White}, N.~E. and {Barthelmy}, S.~D. and {Burrows}, D.~N. and {Cominsky}, L.~R. and {Hurley}, K.~C. and {Marshall}, F.~E. and {M{\'e}sz{\'a}ros}, P. and {Roming}, P.~W.~A. and {Angelini}, L. and {Barbier}, L.~M. and {Belloni}, T. and {Campana}, S. and {Caraveo}, P.~A. and {Chester}, M.~M. and {Citterio}, O. and {Cline}, T.~L. and {Cropper}, M.~S. and {Cummings}, J.~R. and {Dean}, A.~J. and {Feigelson}, E.~D. and {Fenimore}, E.~E. and {Frail}, D.~A. and {Fruchter}, A.~S. and {Garmire}, G.~P. and {Gendreau}, K. and {Ghisellini}, G. and {Greiner}, J. and {Hill}, J.~E. and {Hunsberger}, S.~D. and {Krimm}, H.~A. and {Kulkarni}, S.~R. and {Kumar}, P. and {Lebrun}, F. and {Lloyd-Ronning}, N.~M. and {Markwardt}, C.~B. and {Mattson}, B.~J. and {Mushotzky}, R.~F. and {Norris}, J.~P. and {Osborne}, J. and {Paczynski}, B. and {Palmer}, D.~M. and {Park}, H. -S. and {Parsons}, A.~M. and {Paul}, J. and {Rees}, M.~J. and {Reynolds}, C.~S. and {Rhoads}, J.~E. and {Sasseen}, T.~P. and {Schaefer}, B.~E. and {Short}, A.~T. and {Smale}, A.~P. and {Smith}, I.~A. and {Stella}, L. and {Tagliaferri}, G. and {Takahashi}, T. and {Tashiro}, M. and {Townsley}, L.~K. and {Tueller}, J. and {Turner}, M.~J.~L. and {Vietri}, M. and {Voges}, W. and {Ward}, M.~J. and {Willingale}, R. and {Zerbi}, F.~M. and {Zhang}, W.~W.},
        title = "{The Swift Gamma-Ray Burst Mission}",
      journal = {\apj},
         year = 2004,
        month = aug,
       volume = {611},
       number = {2},
        pages = {1005-1020},
          doi = {10.1086/422091},
archivePrefix = {arXiv},
       eprint = {astro-ph/0405233},
 primaryClass = {astro-ph},
       adsurl = {https://ui.adsabs.harvard.edu/abs/2004ApJ...611.1005G}
}

@ARTICLE{XRISM2020arXiv200304962X,
       author = {{XRISM Science Team}},
        title = "{Science with the X-ray Imaging and Spectroscopy Mission (XRISM)}",
      journal = {arXiv e-prints},
         year = 2020,
        month = mar,
          eid = {arXiv:2003.04962},
        pages = {arXiv:2003.04962},
          doi = {10.48550/arXiv.2003.04962},
archivePrefix = {arXiv},
       eprint = {2003.04962},
 primaryClass = {astro-ph.HE},
       adsurl = {https://ui.adsabs.harvard.edu/abs/2020arXiv200304962X}
}

@ARTICLE{Soffitta2021AJ....162..208S,
       author = {{Soffitta}, Paolo and {Baldini}, Luca and {Bellazzini}, Ronaldo and {Costa}, Enrico and {Latronico}, Luca and {Muleri}, Fabio and {Del Monte}, Ettore and {Fabiani}, Sergio and {Minuti}, Massimo and {Pinchera}, Michele and {Sgro'}, Carmelo and {Spandre}, Gloria and {Trois}, Alessio and {Amici}, Fabrizio and {Andersson}, Hans and {Attina'}, Primo and {Bachetti}, Matteo and {Barbanera}, Mattia and {Borotto}, Fabio and {Brez}, Alessandro and {Brienza}, Daniele and {Caporale}, Ciro and {Cardelli}, Claudia and {Carpentiero}, Rita and {Castellano}, Simone and {Castronuovo}, Marco and {Cavalli}, Luca and {Cavazzuti}, Elisabetta and {Ceccanti}, Marco and {Centrone}, Mauro and {Ciprini}, Stefano and {Citraro}, Saverio and {D'Amico}, Fabio and {D'Alba}, Elisa and {Di Cosimo}, Sergio and {Di Lalla}, Niccolo' and {Di Marco}, Alessandro and {Di Persio}, Giuseppe and {Donnarumma}, Immacolata and {Evangelista}, Yuri and {Ferrazzoli}, Riccardo and {Hayato}, Asami and {Kitaguchi}, Takao and {La Monaca}, Fabio and {Lefevre}, Carlo and {Loffredo}, Pasqualino and {Lorenzi}, Paolo and {Lucchesi}, Leonardo and {Magazzu}, Carlo and {Maldera}, Simone and {Manfreda}, Alberto and {Mangraviti}, Elio and {Marengo}, Marco and {Matt}, Giorgio and {Mereu}, Paolo and {Morbidini}, Alfredo and {Mosti}, Federico and {Nakano}, Toshio and {Nasimi}, Hikmat and {Negri}, Barbara and {Nenonen}, Seppo and {Nuti}, Alessio and {Orsini}, Leonardo and {Perri}, Matteo and {Pesce-Rollins}, Melissa and {Piazzolla}, Raffaele and {Pilia}, Maura and {Profeti}, Alessandro and {Puccetti}, Simonetta and {Rankin}, John and {Ratheesh}, Ajay and {Rubini}, Alda and {Santoli}, Francesco and {Sarra}, Paolo and {Scalise}, Emanuele and {Sciortino}, Andrea and {Tamagawa}, Toru and {Tardiola}, Marcello and {Tobia}, Antonino and {Vimercati}, Marco and {Xie}, Fei},
        title = "{The Instrument of the Imaging X-Ray Polarimetry Explorer}",
      journal = {\aj},
         year = 2021,
        month = nov,
       volume = {162},
       number = {5},
          eid = {208},
        pages = {208},
          doi = {10.3847/1538-3881/ac19b0},
archivePrefix = {arXiv},
       eprint = {2108.00284},
 primaryClass = {astro-ph.IM},
       adsurl = {https://ui.adsabs.harvard.edu/abs/2021AJ....162..208S}
}

@ARTICLE{Predehl2021A&A...647A...1P,
       author = {{Predehl}, P. and {Andritschke}, R. and {Arefiev}, V. and {Babyshkin}, V. and {Batanov}, O. and {Becker}, W. and {B{\"o}hringer}, H. and {Bogomolov}, A. and {Boller}, T. and {Borm}, K. and {Bornemann}, W. and {Br{\"a}uninger}, H. and {Br{\"u}ggen}, M. and {Brunner}, H. and {Brusa}, M. and {Bulbul}, E. and {Buntov}, M. and {Burwitz}, V. and {Burkert}, W. and {Clerc}, N. and {Churazov}, E. and {Coutinho}, D. and {Dauser}, T. and {Dennerl}, K. and {Doroshenko}, V. and {Eder}, J. and {Emberger}, V. and {Eraerds}, T. and {Finoguenov}, A. and {Freyberg}, M. and {Friedrich}, P. and {Friedrich}, S. and {F{\"u}rmetz}, M. and {Georgakakis}, A. and {Gilfanov}, M. and {Granato}, S. and {Grossberger}, C. and {Gueguen}, A. and {Gureev}, P. and {Haberl}, F. and {H{\"a}lker}, O. and {Hartner}, G. and {Hasinger}, G. and {Huber}, H. and {Ji}, L. and {Kienlin}, A. v. and {Kink}, W. and {Korotkov}, F. and {Kreykenbohm}, I. and {Lamer}, G. and {Lomakin}, I. and {Lapshov}, I. and {Liu}, T. and {Maitra}, C. and {Meidinger}, N. and {Menz}, B. and {Merloni}, A. and {Mernik}, T. and {Mican}, B. and {Mohr}, J. and {M{\"u}ller}, S. and {Nandra}, K. and {Nazarov}, V. and {Pacaud}, F. and {Pavlinsky}, M. and {Perinati}, E. and {Pfeffermann}, E. and {Pietschner}, D. and {Ramos-Ceja}, M.~E. and {Rau}, A. and {Reiffers}, J. and {Reiprich}, T.~H. and {Robrade}, J. and {Salvato}, M. and {Sanders}, J. and {Santangelo}, A. and {Sasaki}, M. and {Scheuerle}, H. and {Schmid}, C. and {Schmitt}, J. and {Schwope}, A. and {Shirshakov}, A. and {Steinmetz}, M. and {Stewart}, I. and {Str{\"u}der}, L. and {Sunyaev}, R. and {Tenzer}, C. and {Tiedemann}, L. and {Tr{\"u}mper}, J. and {Voron}, V. and {Weber}, P. and {Wilms}, J. and {Yaroshenko}, V.},
        title = "{The eROSITA X-ray telescope on SRG}",
      journal = {\aap},
         year = 2021,
        month = mar,
       volume = {647},
          eid = {A1},
        pages = {A1},
          doi = {10.1051/0004-6361/202039313},
archivePrefix = {arXiv},
       eprint = {2010.03477},
 primaryClass = {astro-ph.HE},
       adsurl = {https://ui.adsabs.harvard.edu/abs/2021A&A...647A...1P}
}

@INPROCEEDINGS{Gendreau2016SPIE.9905E..1HG,
       author = {{Gendreau}, Keith C. and {Arzoumanian}, Zaven and {Adkins}, Phillip W. and {Albert}, Cheryl L. and {Anders}, John F. and {Aylward}, Andrew T. and {Baker}, Charles L. and {Balsamo}, Erin R. and {Bamford}, William A. and {Benegalrao}, Suyog S. and {Berry}, Daniel L. and {Bhalwani}, Shiraz and {Black}, J. Kevin and {Blaurock}, Carl and {Bronke}, Ginger M. and {Brown}, Gary L. and {Budinoff}, Jason G. and {Cantwell}, Jeffrey D. and {Cazeau}, Thoniel and {Chen}, Philip T. and {Clement}, Thomas G. and {Colangelo}, Andrew T. and {Coleman}, Jerry S. and {Coopersmith}, Jonathan D. and {Dehaven}, William E. and {Doty}, John P. and {Egan}, Mark D. and {Enoto}, Teruaki and {Fan}, Terry W. and {Ferro}, Deneen M. and {Foster}, Richard and {Galassi}, Nicholas M. and {Gallo}, Luis D. and {Green}, Chris M. and {Grosh}, Dave and {Ha}, Kong Q. and {Hasouneh}, Monther A. and {Heefner}, Kristofer B. and {Hestnes}, Phyllis and {Hoge}, Lisa J. and {Jacobs}, Tawanda M. and {J{\o}rgensen}, John L. and {Kaiser}, Michael A. and {Kellogg}, James W. and {Kenyon}, Steven J. and {Koenecke}, Richard G. and {Kozon}, Robert P. and {LaMarr}, Beverly and {Lambertson}, Mike D. and {Larson}, Anne M. and {Lentine}, Steven and {Lewis}, Jesse H. and {Lilly}, Michael G. and {Liu}, Kuochia Alice and {Malonis}, Andrew and {Manthripragada}, Sridhar S. and {Markwardt}, Craig B. and {Matonak}, Bryan D. and {Mcginnis}, Isaac E. and {Miller}, Roger L. and {Mitchell}, Alissa L. and {Mitchell}, Jason W. and {Mohammed}, Jelila S. and {Monroe}, Charles A. and {Montt de Garcia}, Kristina M. and {Mul{\'e}}, Peter D. and {Nagao}, Louis T. and {Ngo}, Son N. and {Norris}, Eric D. and {Norwood}, Dwight A. and {Novotka}, Joseph and {Okajima}, Takashi and {Olsen}, Lawrence G. and {Onyeachu}, Chimaobi O. and {Orosco}, Henry Y. and {Peterson}, Jacqualine R. and {Pevear}, Kristina N. and {Pham}, Karen K. and {Pollard}, Sue E. and {Pope}, John S. and {Powers}, Daniel F. and {Powers}, Charles E. and {Price}, Samuel R. and {Prigozhin}, Gregory Y. and {Ramirez}, Julian B. and {Reid}, Winston J. and {Remillard}, Ronald A. and {Rogstad}, Eric M. and {Rosecrans}, Glenn P. and {Rowe}, John N. and {Sager}, Jennifer A. and {Sanders}, Claude A. and {Savadkin}, Bruce and {Saylor}, Maxine R. and {Schaeffer}, Alexander F. and {Schweiss}, Nancy S. and {Semper}, Sean R. and {Serlemitsos}, Peter J. and {Shackelford}, Larry V. and {Soong}, Yang and {Struebel}, Jonathan and {Vezie}, Michael L. and {Villasenor}, Joel S. and {Winternitz}, Luke B. and {Wofford}, George I. and {Wright}, Michael R. and {Yang}, Mike Y. and {Yu}, Wayne H.},
        title = "{The Neutron star Interior Composition Explorer (NICER): design and development}",
    booktitle = {Space Telescopes and Instrumentation 2016: Ultraviolet to Gamma Ray},
         year = 2016,
       editor = {{den Herder}, Jan-Willem A. and {Takahashi}, Tadayuki and {Bautz}, Marshall},
       series = {Society of Photo-Optical Instrumentation Engineers (SPIE) Conference Series},
       volume = {9905},
        month = jul,
          eid = {99051H},
        pages = {99051H},
          doi = {10.1117/12.2231304},
       adsurl = {https://ui.adsabs.harvard.edu/abs/2016SPIE.9905E..1HG}
}

@ARTICLE{Harrison2013ApJ...770..103H,
       author = {{Harrison}, Fiona A. and {Craig}, William W. and {Christensen}, Finn E. and {Hailey}, Charles J. and {Zhang}, William W. and {Boggs}, Steven E. and {Stern}, Daniel and {Cook}, W. Rick and {Forster}, Karl and {Giommi}, Paolo and {Grefenstette}, Brian W. and {Kim}, Yunjin and {Kitaguchi}, Takao and {Koglin}, Jason E. and {Madsen}, Kristin K. and {Mao}, Peter H. and {Miyasaka}, Hiromasa and {Mori}, Kaya and {Perri}, Matteo and {Pivovaroff}, Michael J. and {Puccetti}, Simonetta and {Rana}, Vikram R. and {Westergaard}, Niels J. and {Willis}, Jason and {Zoglauer}, Andreas and {An}, Hongjun and {Bachetti}, Matteo and {Barri{\`e}re}, Nicolas M. and {Bellm}, Eric C. and {Bhalerao}, Varun and {Brejnholt}, Nicolai F. and {Fuerst}, Felix and {Liebe}, Carl C. and {Markwardt}, Craig B. and {Nynka}, Melania and {Vogel}, Julia K. and {Walton}, Dominic J. and {Wik}, Daniel R. and {Alexander}, David M. and {Cominsky}, Lynn R. and {Hornschemeier}, Ann E. and {Hornstrup}, Allan and {Kaspi}, Victoria M. and {Madejski}, Greg M. and {Matt}, Giorgio and {Molendi}, Silvano and {Smith}, David M. and {Tomsick}, John A. and {Ajello}, Marco and {Ballantyne}, David R. and {Balokovi{\'c}}, Mislav and {Barret}, Didier and {Bauer}, Franz E. and {Blandford}, Roger D. and {Brandt}, W. Niel and {Brenneman}, Laura W. and {Chiang}, James and {Chakrabarty}, Deepto and {Chenevez}, Jerome and {Comastri}, Andrea and {Dufour}, Francois and {Elvis}, Martin and {Fabian}, Andrew C. and {Farrah}, Duncan and {Fryer}, Chris L. and {Gotthelf}, Eric V. and {Grindlay}, Jonathan E. and {Helfand}, David J. and {Krivonos}, Roman and {Meier}, David L. and {Miller}, Jon M. and {Natalucci}, Lorenzo and {Ogle}, Patrick and {Ofek}, Eran O. and {Ptak}, Andrew and {Reynolds}, Stephen P. and {Rigby}, Jane R. and {Tagliaferri}, Gianpiero and {Thorsett}, Stephen E. and {Treister}, Ezequiel and {Urry}, C. Megan},
        title = "{The Nuclear Spectroscopic Telescope Array (NuSTAR) High-energy X-Ray Mission}",
      journal = {\apj},
         year = 2013,
        month = jun,
       volume = {770},
       number = {2},
          eid = {103},
        pages = {103},
          doi = {10.1088/0004-637X/770/2/103},
archivePrefix = {arXiv},
       eprint = {1301.7307},
 primaryClass = {astro-ph.IM},
       adsurl = {https://ui.adsabs.harvard.edu/abs/2013ApJ...770..103H}
}

@ARTICLE{Struder2001A&A...365L..18S,
       author = {{Str{\"u}der}, L. and {Briel}, U. and {Dennerl}, K. and {Hartmann}, R. and {Kendziorra}, E. and {Meidinger}, N. and {Pfeffermann}, E. and {Reppin}, C. and {Aschenbach}, B. and {Bornemann}, W. and {Br{\"a}uninger}, H. and {Burkert}, W. and {Elender}, M. and {Freyberg}, M. and {Haberl}, F. and {Hartner}, G. and {Heuschmann}, F. and {Hippmann}, H. and {Kastelic}, E. and {Kemmer}, S. and {Kettenring}, G. and {Kink}, W. and {Krause}, N. and {M{\"u}ller}, S. and {Oppitz}, A. and {Pietsch}, W. and {Popp}, M. and {Predehl}, P. and {Read}, A. and {Stephan}, K.~H. and {St{\"o}tter}, D. and {Tr{\"u}mper}, J. and {Holl}, P. and {Kemmer}, J. and {Soltau}, H. and {St{\"o}tter}, R. and {Weber}, U. and {Weichert}, U. and {von Zanthier}, C. and {Carathanassis}, D. and {Lutz}, G. and {Richter}, R.~H. and {Solc}, P. and {B{\"o}ttcher}, H. and {Kuster}, M. and {Staubert}, R. and {Abbey}, A. and {Holland}, A. and {Turner}, M. and {Balasini}, M. and {Bignami}, G.~F. and {La Palombara}, N. and {Villa}, G. and {Buttler}, W. and {Gianini}, F. and {Lain{\'e}}, R. and {Lumb}, D. and {Dhez}, P.},
        title = "{The European Photon Imaging Camera on XMM-Newton: The pn-CCD camera}",
      journal = {\aap},
         year = 2001,
        month = jan,
       volume = {365},
        pages = {L18-L26},
          doi = {10.1051/0004-6361:20000066},
       adsurl = {https://ui.adsabs.harvard.edu/abs/2001A&A...365L..18S}
}

@ARTICLE{Jiang2025arXiv250509671J,
       author = {{Jiang}, Yan-Fei and {Blaes}, Omer and {Kaul}, Ish and {Zhang}, Lizhong},
        title = "{Radiation and Magnetic Pressure Support in Accretion Disks around Supermassive Black Holes and The Physical Origin of the Extreme Ultraviolet to Soft X-ray Spectrum}",
      journal = {arXiv e-prints},
         year = 2025,
        month = may,
          eid = {arXiv:2505.09671},
        pages = {arXiv:2505.09671},
          doi = {10.48550/arXiv.2505.09671},
archivePrefix = {arXiv},
       eprint = {2505.09671},
 primaryClass = {astro-ph.HE},
       adsurl = {https://ui.adsabs.harvard.edu/abs/2025arXiv250509671J}
}

@ARTICLE{Rozanska2015A&A...580A..77R,
       author = {{R{\'o}{\.z}a{\'n}ska}, A. and {Malzac}, J. and {Belmont}, R. and {Czerny}, B. and {Petrucci}, P. -O.},
        title = "{Warm and optically thick dissipative coronae above accretion disks}",
      journal = {\aap},
         year = 2015,
        month = aug,
       volume = {580},
          eid = {A77},
        pages = {A77},
          doi = {10.1051/0004-6361/201526288},
archivePrefix = {arXiv},
       eprint = {1504.03160},
 primaryClass = {astro-ph.GA},
       adsurl = {https://ui.adsabs.harvard.edu/abs/2015A&A...580A..77R}
}

@ARTICLE{Kawashima2023ApJ...949..101K,
       author = {{Kawashima}, Tomohisa and {Ohsuga}, Ken and {Takahashi}, Hiroyuki R.},
        title = "{RAIKOU (来光): A General Relativistic, Multiwavelength Radiative Transfer Code}",
      journal = {\apj},
         year = 2023,
        month = jun,
       volume = {949},
       number = {2},
          eid = {101},
        pages = {101},
          doi = {10.3847/1538-4357/acc94a},
archivePrefix = {arXiv},
       eprint = {2108.05131},
 primaryClass = {astro-ph.HE},
       adsurl = {https://ui.adsabs.harvard.edu/abs/2023ApJ...949..101K}
}

@ARTICLE{Narayan2016MNRAS.457..608N,
       author = {{Narayan}, Ramesh and {Zhu}, Yucong and {Psaltis}, Dimitrios and {Sadowski}, Aleksander},
        title = "{HEROIC: 3D general relativistic radiative post-processor with comptonization for black hole accretion discs}",
      journal = {\mnras},
         year = 2016,
        month = mar,
       volume = {457},
       number = {1},
        pages = {608-628},
          doi = {10.1093/mnras/stv2979},
archivePrefix = {arXiv},
       eprint = {1510.04208},
 primaryClass = {astro-ph.HE},
       adsurl = {https://ui.adsabs.harvard.edu/abs/2016MNRAS.457..608N}
}

@ARTICLE{Mills2024ApJ...974..166M,
       author = {{Mills}, Brianna S. and {Davis}, Shane W. and {Jiang}, Yan-Fei and {Middleton}, Matthew J.},
        title = "{Spectral Calculations of 3D Radiation Magnetohydrodynamic Simulations of Super-Eddington Accretion onto a Stellar-mass Black Hole}",
      journal = {\apj},
         year = 2024,
        month = oct,
       volume = {974},
       number = {2},
          eid = {166},
        pages = {166},
          doi = {10.3847/1538-4357/ad6b21},
archivePrefix = {arXiv},
       eprint = {2304.07977},
 primaryClass = {astro-ph.HE},
       adsurl = {https://ui.adsabs.harvard.edu/abs/2024ApJ...974..166M}
}

@ARTICLE{Zhang2019ApJ...875..148Z,
       author = {{Zhang}, Wenda and {Dov{\v{c}}iak}, Michal and {Bursa}, Michal},
        title = "{Constraining the Size of the Corona with Fully Relativistic Calculations of Spectra of Extended Coronae. I. The Monte Carlo Radiative Transfer Code}",
      journal = {\apj},
         year = 2019,
        month = apr,
       volume = {875},
       number = {2},
          eid = {148},
        pages = {148},
          doi = {10.3847/1538-4357/ab1261},
archivePrefix = {arXiv},
       eprint = {1903.09241},
 primaryClass = {astro-ph.HE},
       adsurl = {https://ui.adsabs.harvard.edu/abs/2019ApJ...875..148Z}
}

@ARTICLE{Dolence2009ApJS..184..387D,
       author = {{Dolence}, Joshua C. and {Gammie}, Charles F. and {Mo{\'s}cibrodzka}, Monika and {Leung}, Po Kin},
        title = "{grmonty: A Monte Carlo Code for Relativistic Radiative Transport}",
      journal = {\apjs},
         year = 2009,
        month = oct,
       volume = {184},
       number = {2},
        pages = {387-397},
          doi = {10.1088/0067-0049/184/2/387},
archivePrefix = {arXiv},
       eprint = {0909.0708},
 primaryClass = {astro-ph.HE},
       adsurl = {https://ui.adsabs.harvard.edu/abs/2009ApJS..184..387D}
}

@ARTICLE{Mihalas1985JCoPh..57....1M,
       author = {{Mihalas}, D.},
        title = "{The computation of radiation transport using Feautrier variables. I - Static media}",
      journal = {Journal of Computational Physics},
         year = 1985,
        month = jan,
       volume = {57},
        pages = {1-25},
          doi = {10.1016/0021-9991(85)90050-6},
       adsurl = {https://ui.adsabs.harvard.edu/abs/1985JCoPh..57....1M}
}

@ARTICLE{Shakura1973A&A....24..337S,
       author = {{Shakura}, N.~I. and {Sunyaev}, R.~A.},
        title = "{Black holes in binary systems. Observational appearance.}",
      journal = {\aap},
         year = 1973,
        month = jan,
       volume = {24},
        pages = {337-355},
       adsurl = {https://ui.adsabs.harvard.edu/abs/1973A&A....24..337S}
}

@ARTICLE{BHK2008,
       author = {{Beckwith}, Kris and {Hawley}, John F. and {Krolik}, Julian H.},
        title = "{The Influence of Magnetic Field Geometry on the Evolution of Black Hole Accretion Flows: Similar Disks, Drastically Different Jets}",
      journal = {\apj},
         year = 2008,
        month = may,
       volume = {678},
       number = {2},
        pages = {1180-1199},
          doi = {10.1086/533492},
archivePrefix = {arXiv},
       eprint = {0709.3833},
 primaryClass = {astro-ph},
       adsurl = {https://ui.adsabs.harvard.edu/abs/2008ApJ...678.1180B}
}

@ARTICLE{Balbus1998RvMP...70....1B,
       author = {{Balbus}, Steven A. and {Hawley}, John F.},
        title = "{Instability, turbulence, and enhanced transport in accretion disks}",
      journal = {Reviews of Modern Physics},
         year = 1998,
        month = jan,
       volume = {70},
       number = {1},
        pages = {1-53},
          doi = {10.1103/RevModPhys.70.1},
       adsurl = {https://ui.adsabs.harvard.edu/abs/1998RvMP...70....1B}
}

@ARTICLE{Ghez2008ApJ...689.1044G,
       author = {{Ghez}, A.~M. and {Salim}, S. and {Weinberg}, N.~N. and {Lu}, J.~R. and {Do}, T. and {Dunn}, J.~K. and {Matthews}, K. and {Morris}, M.~R. and {Yelda}, S. and {Becklin}, E.~E. and {Kremenek}, T. and {Milosavljevic}, M. and {Naiman}, J.},
        title = "{Measuring Distance and Properties of the Milky Way's Central Supermassive Black Hole with Stellar Orbits}",
      journal = {\apj},
         year = 2008,
        month = dec,
       volume = {689},
       number = {2},
        pages = {1044-1062},
          doi = {10.1086/592738},
archivePrefix = {arXiv},
       eprint = {0808.2870},
 primaryClass = {astro-ph},
       adsurl = {https://ui.adsabs.harvard.edu/abs/2008ApJ...689.1044G}
}

@ARTICLE{Mendoza2021Atoms...9...12M,
       author = {{Mendoza}, Claudio and {Bautista}, Manuel A. and {Deprince}, J{\'e}r{\^o}me and {Garc{\'\i}a}, Javier A. and {Gatuzz}, Efra{\'\i}n and {Gorczyca}, Thomas W. and {Kallman}, Timothy R. and {Palmeri}, Patrick and {Quinet}, Pascal and {Witthoeft}, Michael C.},
        title = "{The XSTAR Atomic Database}",
      journal = {Atoms},
         year = 2021,
        month = feb,
       volume = {9},
       number = {1},
          eid = {12},
        pages = {12},
          doi = {10.3390/atoms9010012},
archivePrefix = {arXiv},
       eprint = {2012.02041},
 primaryClass = {astro-ph.IM},
       adsurl = {https://ui.adsabs.harvard.edu/abs/2021Atoms...9...12M}
}

@ARTICLE{Kallman2001ApJS..133..221K,
       author = {{Kallman}, T. and {Bautista}, M.},
        title = "{Photoionization and High-Density Gas}",
      journal = {\apjs},
         year = 2001,
        month = mar,
       volume = {133},
       number = {1},
        pages = {221-253},
          doi = {10.1086/319184},
       adsurl = {https://ui.adsabs.harvard.edu/abs/2001ApJS..133..221K}
}

@INPROCEEDINGS{Novikov1973blho.conf..343N,
       author = {{Novikov}, I.~D. and {Thorne}, K.~S.},
        title = "{Astrophysics of black holes.}",
    booktitle = {Black Holes (Les Astres Occlus)},
         year = 1973,
       editor = {{Dewitt}, C. and {Dewitt}, B.~S.},
        month = jan,
        pages = {343-450},
       adsurl = {https://ui.adsabs.harvard.edu/abs/1973blho.conf..343N}
}

@ARTICLE{Reynolds2003PhR...377..389R,
       author = {{Reynolds}, Christopher S. and {Nowak}, Michael A.},
        title = "{Fluorescent iron lines as a probe of astrophysical black hole systems}",
      journal = {\physrep},
         year = 2003,
        month = apr,
       volume = {377},
       number = {6},
        pages = {389-466},
          doi = {10.1016/S0370-1573(02)00584-7},
archivePrefix = {arXiv},
       eprint = {astro-ph/0212065},
 primaryClass = {astro-ph},
       adsurl = {https://ui.adsabs.harvard.edu/abs/2003PhR...377..389R}
}

@ARTICLE{DeVilliers2003ApJ...599.1238D,
       author = {{De Villiers}, Jean-Pierre and {Hawley}, John F. and {Krolik}, Julian H.},
        title = "{Magnetically Driven Accretion Flows in the Kerr Metric. I. Models and Overall Structure}",
      journal = {\apj},
         year = 2003,
        month = dec,
       volume = {599},
       number = {2},
        pages = {1238-1253},
          doi = {10.1086/379509},
archivePrefix = {arXiv},
       eprint = {astro-ph/0307260},
 primaryClass = {astro-ph},
       adsurl = {https://ui.adsabs.harvard.edu/abs/2003ApJ...599.1238D}
}

@ARTICLE{Magdziarz1998MNRAS.301..179M,
       author = {{Magdziarz}, Pawel and {Blaes}, Omer M. and {Zdziarski}, Andrzej A. and {Johnson}, W. Neil and {Smith}, David A.},
        title = "{A spectral decomposition of the variable optical, ultraviolet and X-ray continuum of NGC 5548}",
      journal = {\mnras},
         year = 1998,
        month = nov,
       volume = {301},
       number = {1},
        pages = {179-192},
          doi = {10.1046/j.1365-8711.1998.02015.x},
       adsurl = {https://ui.adsabs.harvard.edu/abs/1998MNRAS.301..179M}
}

@ARTICLE{Gierlinski2004MNRAS.349L...7G,
       author = {{Gierli{\'n}ski}, Marek and {Done}, Chris},
        title = "{Is the soft excess in active galactic nuclei real?}",
      journal = {\mnras},
         year = 2004,
        month = mar,
       volume = {349},
       number = {1},
        pages = {L7-L11},
          doi = {10.1111/j.1365-2966.2004.07687.x},
archivePrefix = {arXiv},
       eprint = {astro-ph/0312271},
 primaryClass = {astro-ph},
       adsurl = {https://ui.adsabs.harvard.edu/abs/2004MNRAS.349L...7G}
}

@ARTICLE{Bianchi2009A&A...495..421B,
       author = {{Bianchi}, S. and {Guainazzi}, M. and {Matt}, G. and {Fonseca Bonilla}, N. and {Ponti}, G.},
        title = "{CAIXA: a catalogue of AGN in the XMM-Newton archive. I. Spectral analysis}",
      journal = {\aap},
         year = 2009,
        month = feb,
       volume = {495},
       number = {2},
        pages = {421-430},
          doi = {10.1051/0004-6361:200810620},
archivePrefix = {arXiv},
       eprint = {0811.1126},
 primaryClass = {astro-ph},
       adsurl = {https://ui.adsabs.harvard.edu/abs/2009A&A...495..421B}
}

@ARTICLE{Kormendy2013ARA&A..51..511K,
       author = {{Kormendy}, John and {Ho}, Luis C.},
        title = "{Coevolution (Or Not) of Supermassive Black Holes and Host Galaxies}",
      journal = {\araa},
         year = 2013,
        month = aug,
       volume = {51},
       number = {1},
        pages = {511-653},
          doi = {10.1146/annurev-astro-082708-101811},
archivePrefix = {arXiv},
       eprint = {1304.7762},
 primaryClass = {astro-ph.CO},
       adsurl = {https://ui.adsabs.harvard.edu/abs/2013ARA&A..51..511K}
}

@ARTICLE{Schodel2002Natur.419..694S,
       author = {{Sch{\"o}del}, R. and {Ott}, T. and {Genzel}, R. and {Hofmann}, R. and {Lehnert}, M. and {Eckart}, A. and {Mouawad}, N. and {Alexander}, T. and {Reid}, M.~J. and {Lenzen}, R. and {Hartung}, M. and {Lacombe}, F. and {Rouan}, D. and {Gendron}, E. and {Rousset}, G. and {Lagrange}, A. -M. and {Brandner}, W. and {Ageorges}, N. and {Lidman}, C. and {Moorwood}, A.~F.~M. and {Spyromilio}, J. and {Hubin}, N. and {Menten}, K.~M.},
        title = "{A star in a 15.2-year orbit around the supermassive black hole at the centre of the Milky Way}",
      journal = {\nat},
         year = 2002,
        month = oct,
       volume = {419},
       number = {6908},
        pages = {694-696},
          doi = {10.1038/nature01121},
archivePrefix = {arXiv},
       eprint = {astro-ph/0210426},
 primaryClass = {astro-ph},
       adsurl = {https://ui.adsabs.harvard.edu/abs/2002Natur.419..694S}
}

@ARTICLE{Kormendy1995ARA&A..33..581K,
       author = {{Kormendy}, John and {Richstone}, Douglas},
        title = "{Inward Bound---The Search For Supermassive Black Holes In Galactic Nuclei}",
      journal = {\araa},
         year = 1995,
        month = jan,
       volume = {33},
        pages = {581},
          doi = {10.1146/annurev.aa.33.090195.003053},
       adsurl = {https://ui.adsabs.harvard.edu/abs/1995ARA&A..33..581K}
}

@ARTICLE{Balbus1991ApJ...376..214B,
       author = {{Balbus}, Steven A. and {Hawley}, John F.},
        title = "{A Powerful Local Shear Instability in Weakly Magnetized Disks. I. Linear Analysis}",
      journal = {\apj},
         year = 1991,
        month = jul,
       volume = {376},
        pages = {214},
          doi = {10.1086/170270},
       adsurl = {https://ui.adsabs.harvard.edu/abs/1991ApJ...376..214B}
}

@ARTICLE{Garcia2014ApJ...782...76G,
       author = {{Garc{\'\i}a}, J. and {Dauser}, T. and {Lohfink}, A. and {Kallman}, T.~R. and {Steiner}, J.~F. and {McClintock}, J.~E. and {Brenneman}, L. and {Wilms}, J. and {Eikmann}, W. and {Reynolds}, C.~S. and {Tombesi}, F.},
        title = "{Improved Reflection Models of Black Hole Accretion Disks: Treating the Angular Distribution of X-Rays}",
      journal = {\apj},
         year = 2014,
        month = feb,
       volume = {782},
       number = {2},
          eid = {76},
        pages = {76},
          doi = {10.1088/0004-637X/782/2/76},
archivePrefix = {arXiv},
       eprint = {1312.3231},
 primaryClass = {astro-ph.HE},
       adsurl = {https://ui.adsabs.harvard.edu/abs/2014ApJ...782...76G}
}

@ARTICLE{Garcia2011ApJ...731..131G,
       author = {{Garc{\'\i}a}, J. and {Kallman}, T.~R. and {Mushotzky}, R.~F.},
        title = "{X-ray Reflected Spectra from Accretion Disk Models. II. Diagnostic Tools for X-ray Observations}",
      journal = {\apj},
         year = 2011,
        month = apr,
       volume = {731},
       number = {2},
          eid = {131},
        pages = {131},
          doi = {10.1088/0004-637X/731/2/131},
archivePrefix = {arXiv},
       eprint = {1101.1115},
 primaryClass = {astro-ph.HE},
       adsurl = {https://ui.adsabs.harvard.edu/abs/2011ApJ...731..131G}
}

@ARTICLE{Ross2005MNRAS.358..211R,
       author = {{Ross}, R.~R. and {Fabian}, A.~C.},
        title = "{A comprehensive range of X-ray ionized-reflection models}",
      journal = {\mnras},
         year = 2005,
        month = mar,
       volume = {358},
       number = {1},
        pages = {211-216},
          doi = {10.1111/j.1365-2966.2005.08797.x},
archivePrefix = {arXiv},
       eprint = {astro-ph/0501116},
 primaryClass = {astro-ph},
       adsurl = {https://ui.adsabs.harvard.edu/abs/2005MNRAS.358..211R}
}

@ARTICLE{Liu2024arXiv241201984L,
       author = {{Liu}, Rongrong and {Nagele}, Chris and {Krolik}, Julian H. and {Kinch}, Brooks E. and {Schnittman}, Jeremy D.},
        title = "{Simulation-based Prediction of Black Hole X-Ray Spectra and Spectral Variability}",
      journal = {\apj},
         year = 2025,
        month = apr,
       volume = {982},
       number = {2},
          eid = {128},
        pages = {128},
          doi = {10.3847/1538-4357/adb61f},
archivePrefix = {arXiv},
       eprint = {2412.01984},
 primaryClass = {astro-ph.HE},
       adsurl = {https://ui.adsabs.harvard.edu/abs/2025ApJ...982..128L}
}

@ARTICLE{Roth2022ApJ...933..226R,
       author = {{Roth}, Nathaniel and {Anninos}, Peter and {Robinson}, Peter B. and {Peterson}, J. Luc and {Polak}, Brooke and {Mangan}, Tymothy K. and {Beyer}, Kyle},
        title = "{General Relativistic Implicit Monte Carlo Radiation-hydrodynamics}",
      journal = {\apj},
         year = 2022,
        month = jul,
       volume = {933},
       number = {2},
          eid = {226},
        pages = {226},
          doi = {10.3847/1538-4357/ac75cb},
archivePrefix = {arXiv},
       eprint = {2206.01760},
 primaryClass = {astro-ph.IM},
       adsurl = {https://ui.adsabs.harvard.edu/abs/2022ApJ...933..226R}
}

@ARTICLE{Roth2025arXiv250118040R,
       author = {{Roth}, Nathaniel and {Anninos}, Peter and {Fragile}, P. Chris and {Pickrel}, Derrick},
        title = "{X-ray Spectra from General Relativistic RMHD Simulations of Thin Disks}",
      journal = {arXiv e-prints},
         year = 2025,
        month = jan,
          eid = {arXiv:2501.18040},
        pages = {arXiv:2501.18040},
          doi = {10.48550/arXiv.2501.18040},
archivePrefix = {arXiv},
       eprint = {2501.18040},
 primaryClass = {astro-ph.HE},
       adsurl = {https://ui.adsabs.harvard.edu/abs/2025arXiv250118040R}
}

@ARTICLE{White2023ApJ...949..103W,
       author = {{White}, Christopher J. and {Mullen}, Patrick D. and {Jiang}, Yan-Fei and {Davis}, Shane W. and {Stone}, James M. and {Morozova}, Viktoriya and {Zhang}, Lizhong},
        title = "{An Extension of the Athena++ Code Framework for Radiation-magnetohydrodynamics in General Relativity Using a Finite-solid-angle Discretization}",
      journal = {\apj},
         year = 2023,
        month = jun,
       volume = {949},
       number = {2},
          eid = {103},
        pages = {103},
          doi = {10.3847/1538-4357/acc8cf},
archivePrefix = {arXiv},
       eprint = {2302.04283},
 primaryClass = {astro-ph.HE},
       adsurl = {https://ui.adsabs.harvard.edu/abs/2023ApJ...949..103W}
}

@ARTICLE{Falcke2004A&A...414..895F,
       author = {{Falcke}, H. and {K{\"o}rding}, E. and {Markoff}, S.},
        title = "{A scheme to unify low-power accreting black holes. Jet-dominated accretion flows and the radio/X-ray correlation}",
      journal = {\aap},
         year = 2004,
        month = feb,
       volume = {414},
        pages = {895-903},
          doi = {10.1051/0004-6361:20031683},
archivePrefix = {arXiv},
       eprint = {astro-ph/0305335},
 primaryClass = {astro-ph},
       adsurl = {https://ui.adsabs.harvard.edu/abs/2004A&A...414..895F}
}

@ARTICLE{Moravec2022A&A...662A..28M,
       author = {{Moravec}, Emily and {Svoboda}, Ji{\v{r}}{\'\i} and {Borkar}, Abhijeet and {Boorman}, Peter and {Kynoch}, Daniel and {Panessa}, Francesca and {Mingo}, Beatriz and {Guainazzi}, Matteo},
        title = "{Do radio active galactic nuclei reflect X-ray binary spectral states?}",
      journal = {\aap},
         year = 2022,
        month = jun,
       volume = {662},
          eid = {A28},
        pages = {A28},
          doi = {10.1051/0004-6361/202142870},
archivePrefix = {arXiv},
       eprint = {2202.11116},
 primaryClass = {astro-ph.GA},
       adsurl = {https://ui.adsabs.harvard.edu/abs/2022A&A...662A..28M}
}

@ARTICLE{Ruan2019ApJ...883...76R,
       author = {{Ruan}, John J. and {Anderson}, Scott F. and {Eracleous}, Michael and {Green}, Paul J. and {Haggard}, Daryl and {MacLeod}, Chelsea L. and {Runnoe}, Jessie C. and {Sobolewska}, Malgosia A.},
        title = "{The Analogous Structure of Accretion Flows in Supermassive and Stellar Mass Black Holes: New Insights from Faded Changing-look Quasars}",
      journal = {\apj},
         year = 2019,
        month = sep,
       volume = {883},
       number = {1},
          eid = {76},
        pages = {76},
          doi = {10.3847/1538-4357/ab3c1a},
archivePrefix = {arXiv},
       eprint = {1903.02553},
 primaryClass = {astro-ph.HE},
       adsurl = {https://ui.adsabs.harvard.edu/abs/2019ApJ...883...76R}
}

@article{Kinch_2016,
doi = {10.3847/0004-637X/826/1/52},
url = {https://dx.doi.org/10.3847/0004-637X/826/1/52},
year = {2016},
month = {jul},
publisher = {The American Astronomical Society},
volume = {826},
number = {1},
pages = {52},
author = {Brooks E. Kinch and Jeremy D. Schnittman and Timothy R. Kallman and Julian H. Krolik},
title = {Fe Kα PROFILES FROM SIMULATIONS OF ACCRETING BLACK HOLES},
journal = {The Astrophysical Journal}
}

@article{Kinch_2019,
doi = {10.3847/1538-4357/ab05d5},
url = {https://dx.doi.org/10.3847/1538-4357/ab05d5},
year = {2019},
month = {mar},
publisher = {The American Astronomical Society},
volume = {873},
number = {1},
pages = {71},
author = {Brooks E. Kinch and Jeremy D. Schnittman and Timothy R. Kallman and Julian H. Krolik},
title = {Predicting the X-Ray Spectra of Stellar-mass Black Holes from Simulations},
journal = {The Astrophysical Journal}
}

@article{Kinch_2020,
doi = {10.3847/1538-4357/abc176},
url = {https://dx.doi.org/10.3847/1538-4357/abc176},
year = {2020},
month = {nov},
publisher = {The American Astronomical Society},
volume = {904},
number = {2},
pages = {117},
author = {Brooks E. Kinch and Scott C. Noble and Jeremy D. Schnittman and Julian H. Krolik},
title = {Inverse Compton Cooling in the Coronae of Simulated Black Hole Accretion Flows},
journal = {The Astrophysical Journal}
}

@article{Kinch_2021,
doi = {10.3847/1538-4357/ac2b9a},
url = {https://dx.doi.org/10.3847/1538-4357/ac2b9a},
year = {2021},
month = {dec},
publisher = {The American Astronomical Society},
volume = {922},
number = {2},
pages = {270},
author = {Brooks E. Kinch and Jeremy D. Schnittman and Scott C. Noble and Timothy R. Kallman and Julian H. Krolik},
title = {Spin and Accretion Rate Dependence of Black Hole X-Ray Spectra},
journal = {The Astrophysical Journal}
}

@article{Remillard_2006,
   title={X-Ray Properties of Black-Hole Binaries},
   volume={44},
   ISSN={1545-4282},
   url={http://dx.doi.org/10.1146/annurev.astro.44.051905.092532},
   DOI={10.1146/annurev.astro.44.051905.092532},
   number={1},
   journal={Annual Review of Astronomy and Astrophysics},
   publisher={Annual Reviews},
   author={Remillard, Ronald A. and McClintock, Jeffrey E.},
   year={2006},
   month=sep, pages={49–92} }

@ARTICLE{Noble_2009,
       author = {{Noble}, Scott C. and {Krolik}, Julian H. and {Hawley}, John F.},
        title = "{Direct Calculation of the Radiative Efficiency of an Accretion Disk Around a Black Hole}",
      journal = {\apj},
         year = 2009,
        month = feb,
       volume = {692},
       number = {1},
        pages = {411-421},
          doi = {10.1088/0004-637X/692/1/411},
archivePrefix = {arXiv},
       eprint = {0808.3140},
 primaryClass = {astro-ph},
       adsurl = {https://ui.adsabs.harvard.edu/abs/2009ApJ...692..411N}
}

@ARTICLE{Noble_2010,
       author = {{Noble}, Scott C. and {Krolik}, Julian H. and {Hawley}, John F.},
        title = "{Dependence of Inner Accretion Disk Stress on Parameters: The Schwarzschild Case}",
      journal = {\apj},
         year = 2010,
        month = mar,
       volume = {711},
       number = {2},
        pages = {959-973},
          doi = {10.1088/0004-637X/711/2/959},
archivePrefix = {arXiv},
       eprint = {1001.4809},
 primaryClass = {astro-ph.HE},
       adsurl = {https://ui.adsabs.harvard.edu/abs/2010ApJ...711..959N}
}

@article{Schnittman_2013,
doi = {10.1088/0004-637X/769/2/156},
url = {https://dx.doi.org/10.1088/0004-637X/769/2/156},
year = {2013},
month = {may},
publisher = {The American Astronomical Society},
volume = {769},
number = {2},
pages = {156},
author = {Jeremy D. Schnittman and Julian H. Krolik and Scott C. Noble},
title = {X-RAY SPECTRA FROM MAGNETOHYDRODYNAMIC SIMULATIONS OF ACCRETING BLACK HOLES},
journal = {The Astrophysical Journal}
}

@ARTICLE{Schnittman_2013b,
       author = {{Schnittman}, Jeremy D. and {Krolik}, Julian H.},
        title = "{A Monte Carlo Code for Relativistic Radiation Transport around Kerr Black Holes}",
      journal = {\apj},
         year = 2013,
        month = nov,
       volume = {777},
       number = {1},
          eid = {11},
        pages = {11},
          doi = {10.1088/0004-637X/777/1/11},
archivePrefix = {arXiv},
       eprint = {1302.3214},
 primaryClass = {astro-ph.HE},
       adsurl = {https://ui.adsabs.harvard.edu/abs/2013ApJ...777...11S}
}

@BOOK{Krolik1999,
       author = {{Krolik}, Julian H.},
        title = "{Active galactic nuclei : from the central black hole to the galactic environment}",
         year = 1999,
         publisher = {Princeton University Press},
       adsurl = {https://ui.adsabs.harvard.edu/abs/1999agnc.book.....K}
}

@ARTICLE{Comptt1994,
       author = {{Titarchuk}, Lev},
        title = "{Generalized Comptonization Models and Application to the Recent High-Energy Observations}",
      journal = {\apj},
         year = 1994,
        month = oct,
       volume = {434},
        pages = {570},
          doi = {10.1086/174760},
       adsurl = {https://ui.adsabs.harvard.edu/abs/1994ApJ...434..570T}
}

@ARTICLE{Fragile+2023,
       author = {{Fragile}, P. Chris and {Anninos}, Peter and {Roth}, Nathaniel and {Mishra}, Bhupendra},
        title = "{Multifrequency General Relativistic Radiation Magnetohydrodynamic Simulations of Thin Disks}",
      journal = {\apj},
         year = 2023,
        month = dec,
       volume = {959},
       number = {1},
          eid = {59},
        pages = {59},
          doi = {10.3847/1538-4357/ad096b},
archivePrefix = {arXiv},
       eprint = {2311.00028},
 primaryClass = {astro-ph.HE},
       adsurl = {https://ui.adsabs.harvard.edu/abs/2023ApJ...959...59F}
}

@ARTICLE{Done+2007,
       author = {{Done}, Chris and {Gierli{\'n}ski}, Marek and {Kubota}, Aya},
        title = "{Modelling the behaviour of accretion flows in X-ray binaries. Everything you always wanted to know about accretion but were afraid to ask}",
      journal = {\aapr},
         year = 2007,
        month = dec,
       volume = {15},
       number = {1},
        pages = {1-66},
          doi = {10.1007/s00159-007-0006-1},
archivePrefix = {arXiv},
       eprint = {0708.0148},
 primaryClass = {astro-ph},
       adsurl = {https://ui.adsabs.harvard.edu/abs/2007A&ARv..15....1D}
}

@ARTICLE{Crummy+2006,
       author = {{Crummy}, J. and {Fabian}, A.~C. and {Gallo}, L. and {Ross}, R.~R.},
        title = "{An explanation for the soft X-ray excess in active galactic nuclei}",
      journal = {\mnras},
         year = 2006,
        month = feb,
       volume = {365},
       number = {4},
        pages = {1067-1081},
          doi = {10.1111/j.1365-2966.2005.09844.x},
archivePrefix = {arXiv},
       eprint = {astro-ph/0511457},
 primaryClass = {astro-ph},
       adsurl = {https://ui.adsabs.harvard.edu/abs/2006MNRAS.365.1067C}
}

@ARTICLE{Zhu+2012,
       author = {{Zhu}, Yucong and {Davis}, Shane W. and {Narayan}, Ramesh and {Kulkarni}, Akshay K. and {Penna}, Robert F. and {McClintock}, Jeffrey E.},
        title = "{The eye of the storm: light from the inner plunging region of black hole accretion discs}",
      journal = {\mnras},
         year = 2012,
        month = aug,
       volume = {424},
       number = {4},
        pages = {2504-2521},
          doi = {10.1111/j.1365-2966.2012.21181.x},
archivePrefix = {arXiv},
       eprint = {1202.1530},
 primaryClass = {astro-ph.HE},
       adsurl = {https://ui.adsabs.harvard.edu/abs/2012MNRAS.424.2504Z}
}

@ARTICLE{Hubeny1995,
       author = {{Hubeny}, I. and {Lanz}, T.},
        title = "{Non-LTE Line-blanketed Model Atmospheres of Hot Stars. I. Hybrid Complete Linearization/Accelerated Lambda Iteration Method}",
      journal = {\apj},
         year = 1995,
        month = feb,
       volume = {439},
        pages = {875},
          doi = {10.1086/175226},
       adsurl = {https://ui.adsabs.harvard.edu/abs/1995ApJ...439..875H}
}

@ARTICLE{Kulkarni+2011,
       author = {{Kulkarni}, Akshay K. and {Penna}, Robert F. and {Shcherbakov}, Roman V. and {Steiner}, James F. and {Narayan}, Ramesh and {S{\"a} Dowski}, Aleksander and {Zhu}, Yucong and {McClintock}, Jeffrey E. and {Davis}, Shane W. and {McKinney}, Jonathan C.},
        title = "{Measuring black hole spin by the continuum-fitting method: effect of deviations from the Novikov-Thorne disc model}",
      journal = {\mnras},
         year = 2011,
        month = jun,
       volume = {414},
       number = {2},
        pages = {1183-1194},
          doi = {10.1111/j.1365-2966.2011.18446.x},
archivePrefix = {arXiv},
       eprint = {1102.0010},
 primaryClass = {astro-ph.HE},
       adsurl = {https://ui.adsabs.harvard.edu/abs/2011MNRAS.414.1183K}
}
\bibliographystyle{aasjournal}

\begin{center}  \label{Sec:ack}
    \textbf{Appendix A: A survey of radiation transport simulations for the post-processing GRMHD simulations }
\end{center}

There are many 
GR ray tracing codes that can post-process GRMHD simulations to produce spectra. The vast majority of these are designed with optically thin systems in mind, such as those imaged by the Event Horizon Telescope \citep{Dolence2009ApJS..184..387D,Kawashima2023ApJ...949..101K}. Once the accretion flow becomes strong enough that optically thick regions of the system coalesce, then the problem becomes numerically more challenging whether one uses Monte-Carlo or characteristics methods. 

For higher accretion rate systems, some assumptions must be made about the nature of the radiation coming from the optically thick region. The usual assumption is that the disk is in local thermal equilibrium and is radiating at a temperature determined by classical disk theory (perhaps with a hardening factor) \citep{Schnittman_2013b,Zhang2019ApJ...875..148Z}. These simulations are limited by the assumption that the disk emits a hardened blackbody. In contrast, \citet{Mills2024ApJ...974..166M} were able to run Monte-Carlo radiation transfer calculations within an optically thick disk; they accomplished this by launching a large number of photons ($10^8$) and considering only free-free and Compton interactions (and mostly focussed on azimuthally-averaged data). They do not solve for the temperature iteratively as we do, but because they are post-processing Athena++ RMHD simulations, the gas temperature does take radiative pressure into account in the simulation itself. 

\cite{Narayan2016MNRAS.457..608N} calculated Comptonization of photons emitted from matter described by a GRMHD simulation, but with a purely thermal source function and an opacity list restricted to free-free absorption and Compton scattering, with the latter described by the Kompaneets equation approximation (i.e., in each scatter, photons change energy by at most a small fraction of their initial energy). This code has the capability to solve for temperature by equating the dissipation rate from the GRMHD simulation with Compton cooling or the radiation flux divergence, the former of which is similar to our method. However, in the applications presented in \citet{Narayan2016MNRAS.457..608N}, the precise dissipation rate from the GRMHD simulation is approximated, and they scale the dissipation by the simulation density, likely underestimating the heat generated in optically thin regions. \cite{Kulkarni+2011,Zhu+2012} used a very thorough stellar atmosphere code \citep{Hubeny1995}, but assume an arbitrary dissipation rate based on mass density so that there is no corona and consequently no photoionization. 

Recently, \citet{RanjanDatta2025arXiv250910177R} used the MONK code \citep{Zhang2019ApJ...875..148Z} combined with a simple disk model to investigate the reflection fraction. Their approach iterates between the corona and the disk in the same way that our method does, in order to achieve global energy balance. However, their disk model, a semi-infinite electron cloud, is arbitrary, and they neglect all atomic radiation processes.

Another class of models attempts to predict black hole accretion spectra without post-processing GRMHD data. These 'phenomenological models' \citep{Comptt1994,Ross2005MNRAS.358..211R,Garcia2011ApJ...731..131G,Garcia2014ApJ...782...76G,Huang2025arXiv251212728H} trade simplifying assumptions about the nature of the accreting black hole system (structure of the accretion disk, coronal geometry, bulk properties of the fluid, ionization state) for a fast evaluation of the resultant spectrum. These models produce plausible looking spectra, but their utility in discriminating different accretion scenarios is questionable because they have no basis in fluid dynamics. There has also been a recent attempt to use a hybrid approach incorporating some GRMHD information into a calculation based on phenomenological models. \citet{Shashank2025arXiv250702583S} use xillver combined with a photosphere location determined from GRMHD simulations. Although they are able to use the slab density from the simulation for the xillver calculation (they use the average density between $\tau=1$ and $\tau=10$ in the simulation for the xillver density), they still assume most parameters, such as the temperature and the nature of the ionizing flux.

\begin{center}  \label{Sec:ack}
    \textbf{Appendix B: Major changes made to \texttt{Pandurata}/\texttt{PTransX}}
\end{center}

\begin{figure}
\includegraphics[width=\linewidth]{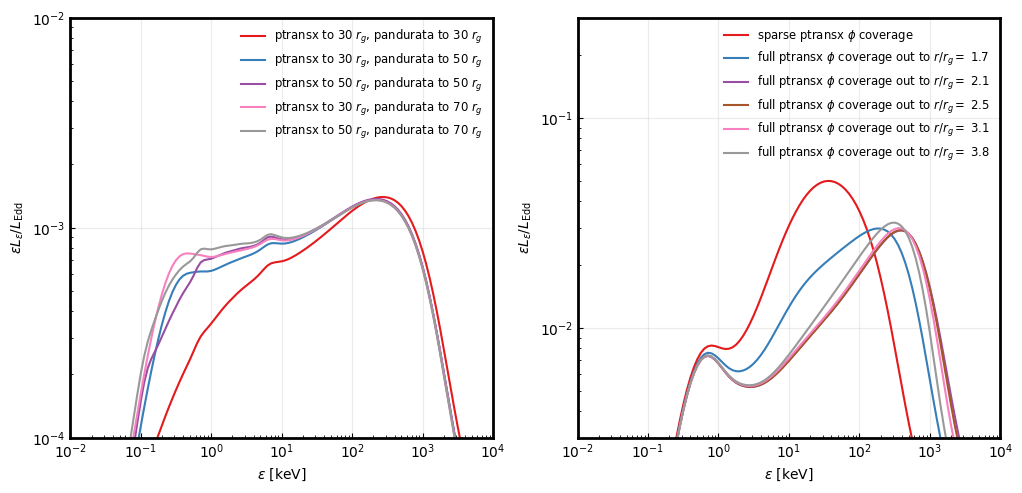}
    \caption{Left panel --- output spectra for the test of the radial coverage of \texttt{Pandurata} and \texttt{PTransX} for the $\mdot=0.01$, $M=10 \;\msun$ model. Right panel --- output spectra for the test of azimuthal coverage of \texttt{PTransX} slabs for the $\mdot=0.1$, $M=10 \;\msun$ model. Note that the first model, labeled sparse, has no additional azimuthal coverage and corresponds to the grid used in previous papers.}
    \label{fig:extant}
\end{figure}

\begin{enumerate}

    \item \textbf{Increased \texttt{PTransX} azimuthal resolution.}

The principal computational cost of our post-processing procedure is the \texttt{PTransX} calculation, which must be performed for many slabs. This can be parallelized to some degree, but we must keep a certain fraction of the cores free for calls to XSTAR or for Jacobian inversion. Computing a \texttt{PTransX} slab for each of the gridpoints in the \texttt{Pandurata} azimuthal and radial grid would be prohibitively expensive. \citet{Kinch_2021} resolved this issue by using a coarser (factor of 3 in $r$, 8 in $\phi$) grid for \texttt{PTransX}, in which the underlying properties of the slab (density and heating rate) were taken from an individual polar column in \texttt{Pandurata} (that is, after interpolating the \texttt{HARM3D} data to the grid used by \texttt{Pandurata}). 

In Fig. \ref{fig:extant}, second panel, we test the validity of this assumption for the $\mdot=0.1,\;M=10\;\msun$ model by performing a sequence of post-processing calculations which have \texttt{PTransX} slabs at every azimuthal gridpoint out to a certain radius, denoted in the legend. The first of these (labeled sparse coverage) does not contain this increased resolution, and its spectrum is markedly different from the higher resolution solutions. This test clearly demonstrates the need for additional \texttt{PTransX} coverage. 

In determining a strategy to applying our computational resources effectively, we noted two properties of slabs closer to the black hole: a) they are quicker to converge ($\sim5$ minutes relative to $\sim2$ hours at $r=r_{\rm inflow}$) and b) their higher surface brightness has an outsized effect on the final spectrum. These facts suggest a gradual unrefinement of the $\phi$ coverage of \texttt{PTransX} (from full coverage to 'sparse' coverage with increasing radius) may optimize the tradeoff between computational time and accuracy, and this is the strategy we adopt (Fig. \ref{fig:tau}). We cannot show the complete convergence of this method as even a single full resolution \texttt{PTransX} run is beyond our current computational resources. We believe that Fig. \ref{fig:extant} is suggestive of this convergence, while noting that better resolution in the future may lead to more accurate results.

    \item \textbf{Revised sector scheme in \texttt{Pandurata}.}

Previous papers used a sector scheme when solving for the coronal temperature. The sectors were typically a region with sides $3 \times 3 \times 4$~\texttt{Pandurata} cells in the radial, polar, and azimuthal directions.  Within each sector, the radiated power was set to the sum of the individual cell powers, and one temperature prevailed across the sector.
This procedure saved computational time, but introduced several artifacts to the spectrum. One such artifact was the creation of single photon packets with unrealistically high energies by \texttt{Pandurata} cells with more power than the other cells in their sector. To eliminate this artifact, \citet{Liu2024arXiv241201984L} masked sectors with $< 36$ scatterings (one scatter per each cell in the sector)
during the previous \texttt{Pandurata} iteration.
We reexamined this question, and determined that solving for temperatures in individual cells was feasible, more realistic, and not much more expensive in most areas of the \texttt{Pandurata} domain. The exception is the region near to the black hole which caused us to modify the temperature iteration as described in the text (Sec. \ref{sec:methods_Pandurata}).
    
    \item \textbf{\texttt{PTransX} cleaving condition.}

As discussed in Sec. \ref{sec:methods_PTransX}, we have modified the cleaving criterion from depending on total optical depth to depending on thermalization optical depth.
This change is more appropriate to finding the boundary of a thermalized region, and makes a particularly large difference for those slabs having a high opacity to scattering, but not to absorption.
    
    \item \textbf{Cleaved slab interior radiative boundary condition.}

Also as discussed in Sec. \ref{sec:methods_PTransX}, we have modified the internal boundary condition of cleaved slabs so that the magnitude of the net outgoing flux from the thermal core matches the mechanical dissipation inside the core. In previous papers, the boundary condition had been a thermal blackbody spectrum the temperature of which was varied with each \texttt{PTransX} temperature update. In theory, both boundary conditions should enforce energy balance in the slab, but in practice, we have found that the previous boundary condition often ran into numerical errors or found unphysical solutions. The new boundary condition is thus less susceptible to numerical artifacts and gives much faster convergence.

    \item \textbf{Modified \texttt{Pandurata} and \texttt{PTransX} radial coverage.}

Fig. \ref{fig:extant} shows the results of a test where we have run the code for different radial extents of both \texttt{Pandurata} and \texttt{PTransX}. In previous works \citep{Kinch_2019,Liu2024arXiv241201984L}, we used \texttt{PTransX} out to $50 \;r_g$ and \texttt{Pandurata} out to $70\;r_g$. In this paper, however, we wanted to limit the \texttt{PTransX} calculations to the region of the \texttt{HARM3D} simulation which had achieved inflow equilibrium (roughly $r/r_g=30$) because the structure of the disk outside this region will not yet have converged. Then the question becomes how much of the \texttt{HARM3D} domain to include in \texttt{Pandurata}. 
As shown in Fig.~\ref{fig:extant}, including only the corona out to the inflow radius clearly misses much of the outgoing radiation, particularly at lower energy, but the spectrum is largely independent of cutoff radius for values $\geq 50 r_g$.
For simplicity, we decided to include the entire \texttt{HARM3D} domain. 

    \item \textbf{Use of the classical disk temperature outside of $r_{\rm inflow}$}

In the previous item, we discussed one implication of computing the coronal emission out to a larger radius than the disk emission.
In previous work with \texttt{Pandurata/PTransX}, the surface brightness
outside the inflow equilibrium region was
taken from the nearest completed \texttt{PTransX} slab. In some cases, however, this procedure severely overestimates the low energy part of the spectrum.

Instead, outside the inflow equilibrium radius we now use the method described in Sec.~\ref{sec:methods_HARM3D}, in which the surface brightness of the disk has a local blackbody spectrum corresponding to the dissipation inside the disk body when the local accretion rate is the same as the rate in the inner disk. 
Because the great majority of the disk outside the inflow equilibrium radius has a thermal core, our results should be closer to what might be found in disks that achieve inflow equilibrium to larger radii than in our simulation.
These thermal spectra outside of $r=30\;r_g$ can be seen in Figs. \ref{fig:B3}, \ref{fig:B4} for reference.

    \item \textbf{Reworked interpolation of cooling function and density from \texttt{HARM3D} and smoothed cooling function within \texttt{PTransX} slabs.}

The \texttt{HARM3D} cooling function
inside the disk body leads to discontinuous spatial variation because many cells inside the disk body have no cooling at the time of the snapshot.
In our previous papers, we interpolate $r$ and $\theta$ simultaneously, but we have since discovered that first interpolating $r$ and then $\theta$ gives more stable volume integrated quantities. Thus, we apply this approach to the interpolation of the cooling function, as well as the density and velocities of \texttt{HARM3D}.

Since the \texttt{HARM3D} cooling function may be sparsely defined (before and after interpolation), in a given \texttt{PTransX} slab, there will be areas of zero dissipation. When these areas happen to fall in the center of the slab, especially for the high mass models, it sometimes leads to catastrophic cooling, with temperatures plummeting below the range well modeled by XSTAR. In an effort to preempt such cooling, we use a spline method to smooth the cooling function so that it is never zero, but that the total dissipation in the entirety of the slab is unchanged. This method is somewhat effective, but as discussed above once the mass gets above $M=10^8\;\msun$, other techniques will need to be utilized.

    \item \textbf{Fixed \texttt{compy} bug affecting Compton downscattering in \texttt{Pandurata}.}

\texttt{Pandurata} makes use of the \texttt{compy} module to compute the redistribution of photon energies after a photon packet scatters in the corona. Specifically, for a given photon energy and electron temperature (both in the fluid frame), \texttt{Pandurata} requires a probability distribution of outgoing photon energies. To provide this, \texttt{compy} pre-computes the photon cumulative distribution function for a given electron temperature and photon energy (both in the fluid frame). \texttt{compy} samples from the velocities of a thermal electron population, then boosts to the rest frame of the sampled electron and computing the Klein-Nishina scattering cross section in that frame. The results of the computed cross sections are tabulated in the form of a cumulative distribution function and a ratio of post-scattering to pre-scattering photon energy. 

\texttt{Pandurata} uses a more compact version of the \texttt{compy} table because it has coarser energy resolution (20 gridpoints per decade for typical runs). In previous papers, during the conversion of the full \texttt{compy} to the compact table, Compton downscattering had been erroneously zeroed out. We have rebuilt the tables after correcting the relevant routine.

    \item \textbf{Fixed bug in volumetric integral of the cooling function.}

\texttt{Pandurata} requires a volumetric integral of the cooling function in order to relate Eq. \ref{Eq:cgs} to the scattering events computed by the Monte Carlo simulation. In previous papers this volume factor was missing one additive term of the metric determinant. We identified and corrected this mistake. The result of this change is that the peak spectral luminosity is reduced by a few percent.

    \item \textbf{Fixed bug in surface brightness reported by \texttt{PTransX} near the black hole. }

Fig. \ref{fig:tau} shows the \texttt{PTransX} coverage of the disk surface. Although we have improved this coverage relative to previous papers (item 1), there are still some vertical columns which are not solved directly by \texttt{PTransX}. For these columns, the seed photon spectrum required by \texttt{Pandurata} is taken from the closest successful \texttt{PTransX} slab. In previous papers, all columns outside of the innermost \texttt{PTransX} slab were assigned seed photon spectra in this manner, even columns which were optically thin, and should not have been assigned seed photons. Fixing this bug results in a decrease of the observed luminosity by a few percent.

    \item \textbf{Fixed bug in \texttt{Pandurata} which in rare cases caused excessive plunging region scattering.}

Where the accretion flow through the plunging region (i.e., inside the ISCO) is optically thick,
the bulk of the optical depth is confined to a few cells on either side of the midplane. If the photon packet step-sizes are too large, a packet incident on this region may scatter after crossing the midplane, but not register that it has passed through the
disk. If 
the photon backscatters, its energy is, on average, amplified.
Repetitions of this process can result in an exponential increase in the packet energy. In one extreme example of this, we discovered a coronal cell in which the recorded inverse Compton power was $\sim 10^{25}\times$ its neighbors' power. Reducing the packet step-size by a factor of ten in this region solves the problem and generally provides better accuracy, in exchange for a modest increase in computational cost.

    \item \textbf{Fixed bug in reflection fraction.}

In order to describe the bug fix, we must first define the reflection fraction. In addition to the seed photon boundary condition at the disk photosphere, a \texttt{Pandurata} run requires knowledge of what happens to photons that reflect from the disk body without being absorbed, so that they re-enter the \texttt{Pandurata} domain. After a \texttt{PTransX} solution and before the next \texttt{Pandurata} solution, we therefore use an additional Monte-Carlo transfer calculation to compute how disk reflection depends on incident angle and energy \citep{Kinch_2021}. To determine the reflection fraction as a function of energy, at every location on the photosphere, we do a new \texttt{PTransX} solution, but with an incident flux having a spectrum flat in terms of photon number. This solution includes photon energy changes by scattering and absorption/re-emission, so it is the reflection fraction in a total sense; it is not the fraction of incident photons at a given energy which are reflected \textit{at that energy}. 
We then perform a Monte Carlo transfer solution taking the local slab to be plane-parallel to determine how 
the spread of outgoing angles depends on the photons' incidence angle.
The convolution of these two results yields the energy and angle dependent reflection fraction, which determines the angle and spectrum of each photon packet reflection in \texttt{Pandurata}. Note that all of these operations preserve photon number and account for relativistic transformations. 

Fig. \ref{fig:reflection} shows the  reflection fraction as a function of photon energy and averaged
over all the reflecting slabs, including both top and bottom disk surfaces,
for all cases.
As this figure shows, the energy-dependence of the disk reflection fractions for our two cases contrast strongly: the $\mdot=0.01$ case is much more reflective at essentially all photon energies than the $\mdot = 0.1$ case.  For whole slabs, which predominate when $\mdot = 0.01$, the reflection fraction is generally close to unity except for ionization edges, such as the H-like Fe K-edge at $\approx 10$ keV. This is because in order to be photon conserving, the reflection includes not only photons which truly reflect within the disk, but also photons which entered the opposite side of the disk and passed through (with a few Compton scatters on the way). Cleaved slabs, on the other hand, which cover much more of the disk for $\mdot=0.1$, have thermal cores, and these absorb all high energy photons, mostly by photoionization, partly by Compton recoil, so that the reflection fraction is closer to 0.6 for most photon energies (not zero due to radiative recombination), and as low as 0.1 in the ionization edges. We stress that this is a physical effect for slabs with a thermal core because they have \textit{much} higher absorption optical depth than their warmer neighbors.

In previous papers, cleaved slabs double-counted photons, so that a typical reflection fraction was $0.6 \times 2= 1.2$, which was rounded down to one. Now the correct value is used.

\begin{figure}
\centering
 \includegraphics[width=0.5\linewidth]{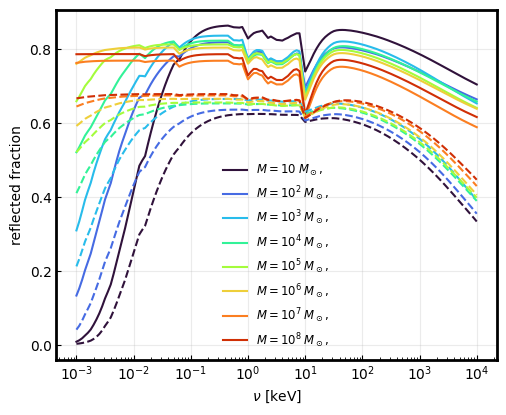}
    \caption{Averaged (hemispherically, radially, azimuthally) photon reflection fraction at a given photon energy for the $\mdot=0.01$ case (solid lines) and the $\mdot=0.1$ case (dashed lines). The photon reflection fraction is the ratio of the outgoing over incoming photon numbers in each energy bin for a flat (in photon number) incoming spectrum. This is not the fraction of photons which are reflected at a given incoming energy (albedo). }
    \label{fig:reflection}
\end{figure}
    
\end{enumerate}

\end{document}